\documentclass{article}%
\usepackage{amsfonts}
\usepackage{amsmath}
\usepackage{color}
\usepackage{isomath}
\usepackage{upgreek}
\usepackage{fancybox}
\usepackage{appendix}%
\usepackage{amssymb}%
\usepackage[dvips]{graphicx}
\usepackage{graphicx}
\usepackage{psfrag}
\usepackage{subcaption}
\usepackage{mathtools}
\usepackage{tikzpagenodes}
\usetikzlibrary{tikzmark}
\usepackage{tikz}
\usetikzlibrary{shapes,snakes}
\usepackage{amsmath,amssymb}
\usepackage{verbatim}
\usepackage{gensymb}

\providecommand{\U}[1]{\protect\rule{.1in}{.1in}}
\newtheorem{theorem}{Theorem}
\newtheorem{acknowledgement}[theorem]{Acknowledgement}

\newtheorem{corollary}[theorem]{Corollary}

\newtheorem{remark}[theorem]{Remark}

\begin{document}

\title{\textbf{Curvature Invariants for accelerating Kerr-Newman black holes in (anti-)de Sitter spacetime}}
\author{G. V. Kraniotis \footnote{email: \textcolor{blue}{gkraniotis@uoi.gr,\;gkraniot@cc.uoi.gr}}\\
University of Ioannina, Physics Department \\ Section of
Theoretical Physics, GR- 451 10, Greece \\
}

 \maketitle

\begin{abstract}
The curvature scalar invariants of the Riemann tensor are important in General Relativity because they allow a manifestly coordinate invariant characterisation of certain geometrical properties of spacetimes such as, among others, curvature singularities, gravitomagnetism. We calculate explicit analytic expressions for the set of Zakhary-McIntosh curvature invariants for accelerating Kerr-Newman black holes in (anti-)de Sitter spacetime as well as for the Kerr-Newman-(anti-)de Sitter black hole.
These black hole metrics belong to the most general type D solution of the Einstein-Maxwell equations with a cosmological constant.
Explicit analytic expressions for the Euler-Poincare density invariant, which is relevant for the computation of the Euler-Poincare characteristic $\chi(M)$, and the Kretschmann scalar are also provided for both cases.
We perform  a detailed plotting of the curvature invariants that reveal a rich structure of the spacetime geometry surrounding the singularity of a rotating, electrically charged and accelerating black hole . These graphs also help us in an exact mathematical way to explore the interior of these black holes.
Our explicit closed form expressions show that the above gravitational backgrounds possess a non-trivial Hirzebruch signature density. Possible physical applications of this property for the electromagnetic duality anomaly in curved spacetimes that can spoil helicity conservation are briefly discussed.

\end{abstract}
\section{Introduction}
The curvature invariants  are important in general relativity since they allow a manifestly coordinate invariant characterisation of certain geometrical properties of spacetimes. In particular they are important in studying curvature singularities \cite{Zahkary}-\cite{Petrov}.
Geometric curvature invariants play a fundamental role in the classification and distinction of exact analytic solutions of General Relativity \cite{LWitten}, particularly for the algebraic Petrov classification \cite{Petrov}  and characterisation of curvature singularities.
They are also significant in discussing the gravitomagnetic properties of spacetimes \cite{Ciufolini},\cite{Filipe}.
To a lesser extent, curvature scalar invariants have also been applied in the field of numerical relativity, chiefly for code testing when evolving exact solutions numerically. They were also considered in the context of black hole collision simulations \cite{BakerJMC}.
Curvature invariants are scalar quantities constructed from tensors that represent curvature. A well known curvature invariant is the Kretschmann scalar which is constructed from the norm of the Riemann curvature tensor $R_{\alpha\beta\gamma\delta}$ \footnote{Kretschmann's scalar originated from the critical examination of Kretschmann on the role of general covariance in the theory of General Relativity \cite{Kretschmann}. Kretschmann recognised that the principle of covariance imposes no restriction on the content of the physical laws, but only on the form in which they are written. }:
\begin{equation}
K:=R_{\alpha\beta\gamma\delta}R^{\alpha\beta\gamma\delta}.
\end{equation}

Carminati and McLenaghan \cite{CarminatiMcL} have shown that more than fourteen independent curvature invariants may be required to describe degenerate cases in the presence of matter. Zahkary and McIntosh (ZM) investigated this problem further and offered the first complete set of seventeen independent real curvature invariants for all possible metrics of Petrov and Segre types \cite{Zahkary}.

It is the purpose of this paper to apply the ZM formalism to the
case of accelerating and rotating charged black holes with non-zero cosmological constant $\Lambda$ and compute for the \textit{first time} analytic algebraic expressions for the corresponding curvature invariants. Specifically, we compute novel closed-form algebraic expressions for the ZM curvature invariants for the
accelerating Kerr-Newman-(anti-)de Sitter black hole.  Using our explicit solutions, we perform an extensive plotting of the curvature invariants that provides a novel pathway to investigate the  spacetime of accelerating Kerr-Newman-(anti-)de Sitter  black holes , its characteristics and explore the interior-a \textit{terra incognita} of these black holes. Moreover, for zero acceleration of the black hole we compute also for  the \textit{first time} exact algebraic expressions for the ZM curvature invariants by applying the ZM-formalism to the Kerr-Newman-(anti-)de Sitter metric.
These black hole metrics belong to the most general type D solution of the Einstein-Maxwell equations with a cosmological constant and constitute the physically most important case \cite{GrifPod},\cite{PlebanskiDemianski}.
Indeed, besides the intrinsic theoretical interest of accelerating or non-accelerating KN(a)dS black holes , a variety of observations support and single out their physical relevance in Nature.

A wide variety of astronomical and cosmological observations in the last two decades, including high-redshift type Ia supernovae, cosmic microwave background radiation and large scale structure indicate convincingly an accelerating expansion of the Universe \cite{Supern},\cite{Jones},\cite{Aubourg},\cite{AbbottTMC}. Such observational data  can be explained by a positive cosmological constant $\Lambda$ ($\Lambda>0$) with a magnitude $\Lambda\sim 10^{-56}{\rm cm}^{-2}$ \cite{GVKSWB}.

 Recent observations of structures near the galactic centre region SgrA*  by the GRAVITY experiment, indicate possible presence of a small electric charge of central supermassive black hole \cite{Zajacek},\cite{Britzen}. Accretion disk physics around magnetised Kerr black holes under the influence of cosmic repulsion  is extensively discussed in the review \cite{universemdpi} \footnote{We also mention that supermassive black holes as possible sources of ultahigh-energy cosmic rays have been suggested in \cite{waldcharge}, where it has been shown that large values of the Lorentz $\gamma$ factor of an escaping ultrahigh-energy particle from the inner regions of the black hole accretion disk may occur only in the presence of the induced charge of the black hole.}.
Therefore, it is quite interesting to study the combined effect of the cosmological constant and electromagnetic fields on the geometry of spacetime surrounding the black hole singularity through the explicit algebraic computation and plotting of the ZM curvature invariants taking into account also the acceleration parameter.

Previous works include the calculation of the Kretschmann scalar for the Kerr-Newman metric \cite{HenryRC}, and the computation for the first two invariants of the ZM set for the Kerr metric in \cite{LakeK}. Expressions for an earlier set of invariants developed in \cite{CarminatiMcL}, have been investigated using, GRTensor, for the Kerr-Newman metric in
\cite{PMusKLak}. The first two ZM invariants as well as the Kretschmann scalar were studied within the Newman-Penrose formalism for the Kerr-Newman metric in \cite{cbcruf}.
Recently, algebraic expressions for the whole set of ZM invariants were produced in \cite{overduincwh} for the Kerr-Newman black hole. Curvature invariants for the accelerating Nat\'{a}rio warp drive were studied in \cite{mattinglyCleav},\cite{GeraldC} while quadratic and some cubic curvature invariants-contractions of the Kummer tensor, were investigated for a Pleba\'{n}ski-Demia\'{n}ski solution in \cite{Boos}.
For a nice general review of the application of symbolic computing to problems in general relativity see \cite{mccallumr}.

The material of this paper is organised as follows: In section \ref{analioteskatazm} we present the definitions of the ZM invariants that we shall use in computing explicit algebraic equations for the curvature invariants for the non-accelerating Kerr-Newman-(anti) de Sitter black hole. In sections \ref{CIZMKNadS} and \ref{CIZMKNadSALL}, we derive  our explicit analytic expressions for the curvature invariants,  Equations (\ref{CIZMein})-(\ref{CIsiebenundzehn}) and (\ref{CHERNPOTRYAGINNOBESCH}) and Theorem \ref{courbureinvriemein}. We also computed explicit algebraic expressions for the topological Euler-Poincare invariant see Eqn.(\ref{LeonardEuler}), and for the Kretschmann scalar see Eqn.(\ref{KretschKerrNewmandeSitter}). In section \ref{RogerEwa} we perform analytical computations within the Newman-Penrose formalism for computing curvature invariants and also derive the syzygies the ZM invariants obey for non-accelerating and accelerating KN(a)dS black holes. In section \ref{Beschleunigung} we first introduce the  type D  metric that solves the Einstein-Maxwell equations with a cosmological constant \cite{PodolskyGrif} (i.e accelerated Kerr-Newman black hole in (anti-)de Sitter spacetime ) and provide a brief discussion of its singularities.  Subsequently we derive \textit{novel} explicit algebraic expressions for the curvature invariants for accelerating Kerr-Newman black holes in (anti-)de Sitter spacetime, eqns. (\ref{AcceBeschlKNdSein})-(\ref{BeschleunKNdSsiebundzehn}) and (\ref{CPBeschleunigungKNdS}), Theorem \ref{courbureinvtheoriazwei}. In section \ref{orizontesgegonotwn} we discuss the horizon radii for accelerating Kerr-Newman black holes with non-zero cosmological constant. Armed with the exact  explicit algebraic expressions for the ZM curvature invariants we derived,  we execute in section \ref{eswterikomelanisopis}, a detailed plotting of these functions for different sets of values of the physical black hole parameters. Finally, we reserve section \ref{symperasmata} for our conclusions and a discussion of the possible physical applications. In particular, we discuss the radiation properties of the curved black hole backgrounds we investigated in this work in relation to photon helicity and quantum chiral anomaly. Both backgrounds possess a non-trivial Chern-Pontryagin  invariant, which is one of the ZM curvature invariants. However for non-accelerating Kerr-Newman-(anti-) de Sitter black holes there is a null contribution to the chiral anomaly since the integrated Chern-Pontryagin-Hirzebruch invariant over all angles is zero, see Eqn.(\ref{nullintegraCPH}). On the other hand, we find that for accelerating rotating and charged black holes with $\Lambda\not =0$, the integrated Chern-Pontryagin-Hirzebruch invariant gives a non-zero result for the quantum photon chiral anomaly, see Eqn.(\ref{nonnullCPHintegraGVK}).


\section{Preliminaries on Riemannian invariants}\label{analioteskatazm}
Taking into account the contribution from the cosmological
constant $\Lambda,$ the generalisation of the Kerr-Newman solution \cite{Newman},\cite{KerrR},
is described by the Kerr-Newman de Sitter $($KNdS$)$ metric
element which in Boyer-Lindquist (BL) coordinates is given by \cite{BCAR},\cite{Stuchlik1},\cite{GrifPod},\cite{ZdeStu} (in units where $G=1$ and $c=1$):
\begin{align}
\mathrm{d}s^{2}  & =-\frac{\Delta_{r}^{KN}}{\Xi^{2}\rho^{2}}(\mathrm{d}%
t-a\sin^{2}\theta\mathrm{d}\phi)^{2}+\frac{\rho^{2}}{\Delta_{r}^{KN}%
}\mathrm{d}r^{2}+\frac{\rho^{2}}{\Delta_{\theta}}\mathrm{d}\theta
^{2}\nonumber \\ &+\frac{\Delta_{\theta}\sin^{2}\theta}{\Xi^{2}\rho^{2}}(a\mathrm{d}%
t-(r^{2}+a^{2})\mathrm{d}\phi)^{2}%
\label{KNADSelement}
\end{align}%
\begin{equation}
\Delta_{\theta}:=1+\frac{a^{2}\Lambda}{3}\cos^{2}\theta,
\;\Xi:=1+\frac {a^{2}\Lambda}{3},
\end{equation}

\begin{equation}
\Delta_{r}^{KN}:=\left(  1-\frac{\Lambda}{3}r^{2}\right)  \left(  r^{2}
+a^{2}\right)  -2mr+q^{2},
\label{DiscrimiL}
\end{equation}

\begin{equation}
\rho^{2}=r^{2}+a^{2}\cos^{2}\theta,
\end{equation}
where $a,m,q,$ denote the Kerr parameter, mass and electric charge
of the black hole, respectively.
The KN(a)dS metric is the most general exact stationary  solution of the Einstein-Maxwell system of differential equations, that represents a non-accelerating, rotating, charged black hole with $\Lambda\not =0$.
This
is accompanied by a non-zero electromagnetic field
$F=\mathrm{d}A,$ where the vector potential is
\cite{GrifPod},\cite{ZST}:
\begin{equation}
A=-\frac{qr}{\Xi(r^{2}+a^{2}\cos^{2}\theta)}(\mathrm{d}t-a\sin^{2}\theta
\mathrm{d}\phi).
\end{equation}

The Christoffel symbols of the second kind are expressed in the coordinate basis in the form:
\begin{equation}
\Gamma^{\lambda}_{\;\mu\nu}=\frac{1}{2}g^{\lambda\alpha}(g_{\mu\alpha,\nu}+g_{\nu\alpha,\mu}-g_{\mu\nu,\alpha}),
\end{equation}
where the summation convention is adopted and a comma denotes a partial derivative.
The Riemann curvature tensor is given by:
\begin{align}
R^{\kappa}_{\;\;\lambda\mu\nu}=\Gamma^{\kappa}_{\;\lambda\nu,\mu}-\Gamma^{\kappa}_{\;\lambda\mu,\nu}+
\Gamma^{\alpha}_{\lambda\nu}\Gamma^{\kappa}_{\;\alpha\mu}-\Gamma^{\alpha}_{\lambda\mu}\Gamma^{\kappa}_{\alpha\nu}.
\end{align}
The symmetric Ricci tensor and the Ricci scalar are defined by:
\begin{equation}
R_{\mu\nu}=R^{\alpha}_{\;\mu\alpha\nu},\;\;\;R=g^{\alpha\beta}R_{\alpha\beta},
\end{equation}
while the Weyl tensor $C_{\kappa\lambda\mu\nu}$ (the trace-free part of the curvature tensor) is given explicitly in terms of the curvature tensor and the metric from the expression:
\begin{align}
C_{\kappa\lambda\mu\nu}=R_{\kappa\lambda\mu\nu}&+\frac{1}{2}(R_{\lambda\mu}g_{\kappa\nu}+R_{\kappa\nu}g_{\lambda\mu}-R_{\lambda\nu}g_{\kappa\mu}-R_{\kappa\mu}g_{\lambda\nu})\nonumber \\&+\frac{1}{6}R(g_{\kappa\mu}g_{\lambda\nu}-g_{\kappa\nu}g_{\lambda\mu}).
\label{WeylH}
\end{align}
The Weyl tensor has in general, ten independent components which at any point are completely independent of the Ricci components. It corresponds to the {\em free gravitational field} \footnote{Globally, however, the Weyl tensor and Ricci tensor are not independent, as they are connected by the differential Bianchi identities. These identities determine the interaction between the free gravitational field and the field sources.} \cite{szekeres}.
The definitions of the Zahkary and McIntosh invariants fall into three groups \cite{Zahkary},\cite{overduincwh}:
\begin{align}
\textit{Weyl invariants}:\;\;\;\;\;&\nonumber\\
&I_1=C_{\alpha\beta\gamma\lambda}C^{\alpha\beta\gamma\lambda}=
C_{\alpha\beta}^{\kappa\lambda}C_{\kappa\lambda}^{\alpha\beta},\\
&I_2=-C_{\alpha\beta}^{\;\;\mu\nu}C_{\mu\nu}^{*\;\;\alpha\beta}=-K_2,\\
&I_3=C_{\alpha\beta}^{\;\;\mu\nu}C_{\mu\nu}^{\;\;o\rho}C_{o\rho}^{\;\;\alpha\beta},\\
&I_4=-C_{\alpha\beta}^{\;\;\kappa\lambda}C_{\kappa\lambda}^{*\;o\rho}C_{o\rho}^{\;\;\alpha\beta}
\end{align}

\begin{align}
\textit{Ricci invariants}:\;\;\;\;\;&\nonumber\\
&I_5=R=g_{\alpha\beta}R^{\alpha\beta},\\
&I_6=R_{\alpha\beta}R^{\alpha\beta}=R_{\alpha\beta}g^{\mu\alpha}g^{\lambda\beta}R_{\mu\lambda},\\
&I_7=R_{\mu}^{\;\nu}R_{\nu}^{\;\rho}R_{\rho}^{\;\mu},\\
&I_8=R_{\mu}^{\;\nu}R_{\nu}^{\;\rho}R_{\rho}^{\;\lambda}R_{\lambda}^{\;\mu}
\end{align}

\begin{align}
\textit{Mixed invariants}:\;\;\;\;\;&\nonumber\\
&I_9=C_{\alpha\beta\mu}^{\;\;\;\;\;\;\nu}R^{\beta\mu}R_{\nu}^{\;\alpha}=-C^{\nu}_{\;\;\mu\alpha\beta}R^{\beta\mu}R_{\nu}^{\;\alpha},\\
&I_{10}=-C^{*\;\;\;\gamma}_{\alpha\beta\lambda}R^{\beta\lambda}R_{\gamma}^{\;\alpha},\\
&I_{11}=R^{\alpha\beta}R^{\mu\nu}(C_{o\alpha\beta}^{\;\;\;\rho}C_{\rho\mu\nu}^{\;\;\;o}-C_{o\alpha\beta}^{*\;\rho}C_{\rho\mu\nu}^{*\;o}),\\
&I_{12}=-R^{\alpha\beta}R^{\mu\nu}(C_{o\alpha\beta}^{*\;\;\;\rho}C_{\rho\mu\nu}^{\;\;\;o}+C_{o\alpha\beta}^{\;\;\;\;\rho}C_{\rho\mu\nu}^{*\;o}),\\
&I_{15}=\frac{1}{16}R^{\alpha\beta}R^{\mu\nu}(C_{o\alpha\beta\rho}C^{o\;\;\;\rho}_{\;\;\mu\nu}+
C^{*}_{o\alpha\beta\rho}C^{*o\;\;\;\rho}_{\;\;\mu\nu}),\\
&I_{16}=-\frac{1}{32}R^{\alpha\beta}R^{\mu\nu}\Biggl(C_{o\eta\sigma\rho}C^{o\;\;\rho}_{\alpha\beta}
C^{\eta\;\;\;\sigma}_{\;\;\mu\nu}+C_{o\eta\sigma\rho}C^{*o\;\;\rho}_{\;\;\alpha\beta}C^{*\eta\;\;\;\sigma}_{\;\;\;\mu\nu}\nonumber \\
&-C^{*}_{o\eta\sigma\rho}C^{*o\;\;\rho}_{\;\;\alpha\beta}C^{\eta\;\;\;\sigma}_{\;\;\mu\nu}+
C^{*}_{o\eta\sigma\rho}C^{o\;\;\rho}_{\alpha\beta}C^{*\eta\;\;\;\sigma}_{\;\;\;\mu\nu}\Biggr),\\
&I_{17}=\frac{1}{32}R^{\alpha\beta}R^{\mu\nu}\Biggl(C^{*}_{o\kappa\lambda\rho}C^{o\;\;\;\rho}_{\;\alpha\beta}
C^{\kappa\;\;\lambda}_{\;\;\mu\nu}+C^{*}_{o\kappa\lambda\rho}C^{*o\;\;\;\rho}_{\;\;\alpha\beta}C^{*\kappa\;\;\;\lambda}_{\;\;\;\mu\nu}\nonumber\\
&-C_{o\kappa\lambda\rho}C^{*o\;\;\;\rho}_{\;\;\alpha\beta}C^{\kappa\;\;\lambda}_{\;\;\mu\nu}+C_{o\kappa\lambda\rho}C^{o\;\;\;\rho}_{\;\alpha\beta}
C^{*\kappa\;\;\;\lambda}_{\;\;\;\mu\nu}\Biggr)
\end{align}
 Here $C_{\alpha\beta\gamma\delta}^{*}$ is the dual of the Weyl tensor, defined by:
 \begin{equation}
 C_{\alpha\beta\gamma\delta}^{*}=\frac{1}{2}E_{\alpha\beta\kappa\lambda}C^{\kappa\lambda}_{\;\;\gamma\delta},
 \end{equation}
 where $E_{\alpha\beta\kappa\lambda}$ is the Levi-Civita pseudotensor.

To calculate the above curvature invariants for the metric (\ref{KNADSelement}) we used Maple\textsuperscript{TM}2021. In section \ref{RogerEwa} we shall use a different method and compute Ricci and Weyl curvature invariants by applying the Newman-Penrose tetrad formalism.
Note that $I_{13}$ and $I_{14}$ both vanish for the  metric (\ref{KNADSelement}), which is of Petrov type D and Segre type [(11)(1,1)] \cite{Zahkary}.

\section{Results for the ZM curvature invariants, the Kretschmann scalar and the Euler invariant for the  Kerr-Newman-(anti-)de Sitter black hole}\label{CIZMKNadS}
We start our computations for the Kerr-Newman-(a)dS black hole with the calculation of the Kretschmann scalar. Indeed, our computations yield:
\begin{theorem}\label{KretschmannKNdS}
The Kretschmann invariant $K=R_{\alpha\beta\gamma\delta}R^{\alpha\beta\gamma\delta}$ for the KN(a)dS black hole is given by the expression:
\begin{align}
&K^{KN(a)dS}=\frac{1}{3\left(r^{2}+a^{2}\cos \! \left(\theta \right)^{2}\right)^{6}}\Biggl(8\Lambda^{2}\cos \! \left(\theta \right)^{12}a^{12}+48\Lambda^{2}\cos \! \left(\theta \right)^{10}a^{10}r^{2}+120\Lambda^{2}\cos \! \left(\theta \right)^{8}a^{8}r^{4}\nonumber \\& +\left(160\Lambda^{2}a^{6}r^{6}-144a^{6}m^{2}\right)\cos \! \left(\theta \right)^{6}+2160\left(\frac{1}{18}r^{8}\Lambda^{2}+m^{2}r^{2}-\frac{2}{3}m \,q^{2}r +\frac{7}{90}q^{4}\right)a^{4}\cos \! \left(\theta \right)^{4}\nonumber \\
&-2160\left(-\frac{1}{45}r^{8}\Lambda^{2}+m^{2}r^{2}-\frac{4}{3}m \,q^{2}r +\frac{17}{45}q^{4}\right)a^{2}r^{2}\cos \! \left(\theta \right)^{2}+8\Lambda^{2}r^{12}+144m^{2}r^{6}-288m \,q^{2}r^{5}+168q^{4}r^{4}\Biggr).
\label{KretschKerrNewmandeSitter}
\end{align}
\end{theorem}
As limiting cases we obtain the following corollaries:
\begin{corollary}
For zero angular momentum of the black hole ($a=0$) i.e. for Reissner-Nordstr\"{o}m-(a)dS black hole the Kretschmann scalar is given by:
\begin{align}
K^{RN(a)dS}=\frac{8\Lambda^{2}r^{12}+144m^{2}r^{6}-288m \,q^{2}r^{5}+168q^{4}r^{4}}{3r^{12}}.
\end{align}
\end{corollary}
\begin{corollary}
For vanishing cosmological constant $\Lambda=0$, i.e for a Kerr-Newman black hole the Kretschmann scalar is given by:
\begin{align}\label{KKN}
&K^{KN}=\frac{1}{\left(r^{2}+a^{2}\cos \! \left(\theta \right)^{2}\right)^{6}}\Biggl(-48\cos \! \left(\theta \right)^{6}a^{6}m^{2}+720\left(m^{2}r^{2}-\frac{2}{3}m \,q^{2}r +\frac{7}{90}q^{4}\right)a^{4}\cos \! \left(\theta \right)^{4}\nonumber \\
&-720\left(m^{2}r^{2}-\frac{4}{3}m \,q^{2}r +\frac{17}{45}q^{4}\right)a^{2}r^{2}\cos \! \left(\theta \right)^{2}+48m^{2}r^{6}-96m \,q^{2}r^{5}+56q^{4}r^{4}\Biggr).
\end{align}
\end{corollary}

Our result  in Corollary (\ref{KKN}), agrees with the result obtained in \cite{HenryRC}.

\begin{corollary}
For zero rotation and zero cosmological constant ($a=0,\Lambda=0$),i.e for a Reissner-Nordstr\"{o}m black hole the Kretschmann invariant is given by:
\begin{equation}
\frac{144m^{2}r^{6}-288m \,q^{2}r^{5}+168q^{4}r^{4}}{3r^{12}}.
\end{equation}
\end{corollary}

 The invariant $I_1=C_{\alpha\beta\gamma\lambda}C^{\alpha\beta\gamma\lambda}$ is related to the Kretschmann scalar $K=R_{\alpha\beta\mu\nu}R^{\alpha\beta\mu\nu}$ the Ricci invariant $I_6$ through the fundamental identity first proven in \cite{Dianyan} :
 \begin{equation}
 C_{\alpha\beta\gamma\lambda}C^{\alpha\beta\gamma\lambda}=R_{\alpha\beta\gamma\lambda}R^{\alpha\beta\gamma\lambda}-2 R_{\alpha\beta}R^{\alpha\beta}+\frac{1}{3}R^2.
 \label{dianyanein}
 \end{equation}
Another non-trivial important identity among the Weyl invariant $I_3$, the Ricci invariants and the Kretschmann scalar is  the following \cite{Dianyan}:
\begin{align}
C_{\alpha\beta}^{\;\;\gamma\sigma}C_{\gamma\sigma}^{\;\;\lambda\tau}C_{\lambda\tau}^{\;\;\alpha\beta}&=
R_{\alpha\beta}^{\;\;\gamma\sigma}R_{\gamma\sigma}^{\;\;\lambda\tau}R_{\lambda\tau}^{\;\;\alpha\beta}-\frac{1}{2}RR_{\alpha\beta\gamma\lambda}R^{\alpha\beta\gamma\lambda}
-6R^{\alpha\gamma}R^{\beta\lambda}R_{\alpha\beta\gamma\lambda}\nonumber\\
&-6R_{\alpha}^{\;\;\beta}R_{\beta}^{\;\;\gamma}R_{\gamma}^{\;\;\alpha}+7RR_{\alpha\beta}R^{\alpha\beta}
-\frac{17}{18}R^3.
\label{dianyanzwei}
\end{align}

 Identities like these, e.g. eqns(\ref{dianyanein})-(\ref{dianyanzwei}), serve as an additional way to test the correctness of our analytic symbolic computations.
 The Chern-Pontryagin invariant $K_2:=C_{\alpha\beta\gamma\delta}^{*}C^{\alpha\beta\gamma\delta}$ is one of the fundamental invariants for the Kerr-Newman-(anti-)de Sitter spacetime.

 We computed analytically this invariant with the following result:

 \begin{align}
K_2= \frac{96a}{\left(r^{2}+a^{2}\cos \! \left(\theta \right)^{2}\right)^{6}} \left(\cos \! \left(\theta \right)^{3}a^{2}m -3\cos \! \left(\theta \right)m \,r^{2}+2\cos \! \left(\theta \right)q^{2}r \right)\nonumber \\
\times \left(-3a^{2}\cos \! \left(\theta \right)^{2}m r +a^{2}\cos \! \left(\theta \right)^{2}q^{2}+m \,r^{3}-q^{2}r^{2}\right).
 \label{CHERNPOTRYAGINNOBESCH}
 \end{align}
  Our result in (\ref{CHERNPOTRYAGINNOBESCH}), agrees with the corresponding expressions obtained in \cite{cbcruf},\cite{overduincwh} for the Kerr-Newman metric.
 We note that the  Chern-Potryagin invariant $K_2$ is proportional to the Kerr parameter $a$ in Equation (\ref{CHERNPOTRYAGINNOBESCH}) and therefore vanishes in the limit of zero angular momentum ($a=0$). Thus, the  Chern-Potryagin invariant characterises the gravitomagnetic properties of the Kerr-Newman-(anti-)de Sitter spacetime. The invariant $K_2$ is also equal to the invariant built from the dual of the Riemmann tensor:
 \begin{equation}
 K_2=C_{\alpha\beta\gamma\delta}^{*}C^{\alpha\beta\gamma\delta}=\frac{1}{2}E^{\alpha\beta\sigma\rho}R_{\sigma\rho}^{\;\mu\nu}
 R_{\alpha\beta\mu\nu}\equiv^{*}\mathbf{R}\cdot\mathbf{R}.
 \label{Hirzebruch}
 \end{equation}

The invariant $^{*}\mathbf{R}\cdot\mathbf{R}=\frac{1}{2}E^{\alpha\beta\sigma\rho}R_{\sigma\rho}^{\;\mu\nu}
 R_{\alpha\beta\mu\nu}$-the Hirzebruch signature density,
 has been proposed by Ciufolini \cite{Ciufolini} to characterise the spacetime geometry and curvature generated by mass-energy currents and by the intrinsic angular momentum of a central body, thus the gravitomagnetic properties of spacetime \footnote{In weak-field general relativity, the angular momentum generated by mass-energy currents plays a role analogous to the magnetic dipole moment of a loop of charge current in Electrodynamics. The electromagnetic analogue of the invariant $^{*}\mathbf{R}\cdot\mathbf{R}$ is of course  $^{*}\mathbf{F}\cdot\mathbf{F}$. However, the latter in electrodynamics characterises the electromagnetic field only, but not the spacetime geometry. While, the former in general relativity characterises the gravitational field and spacetime geometry.}. In \cite{Ciufolini} the  quantity $^{*}\mathbf{R}\cdot\mathbf{R}$ was computed for the Kerr metric. Our result in Eqn.(\ref{CHERNPOTRYAGINNOBESCH}) generalises this computation for the most general Kerr-Newman-(anti-)de Sitter metric. Moreover, in section \ref{Beschleunigung} we will generalise further our result in eqn(\ref{CHERNPOTRYAGINNOBESCH}), by calculating the Chern-Pontryagin invariant $K_2$ for \textit{accelerating} Kerr-Newman black holes in (anti-)de Sitter spacetime, see eqn(\ref{CPBeschleunigungKNdS}).

The Hirzebruch signature density, $K_2$, is of course an example of a topological invariant. Another interesting topological invariant, besides $K_2$, is the quantity constructed from the doubly dual curvature tensor:
\begin{equation}
K_{{\rm Euler}}=\frac{1}{4}E^{\alpha\beta\gamma\delta}E^{\mu\nu\rho\sigma}R_{\alpha\beta\mu\nu}
R_{\gamma\delta\rho\sigma}.
\end{equation}

The topological invariant $K_{{\rm Euler}}$ is essentially Euler's density whose integral over spacetime measure gives the so called Euler-Poincare characteristic $\chi$ \footnote{\label{EulerP}Indeed, the Euler-Poincare characteristic in four dimensions is: $\chi=\int\frac{-1}{128\pi^2}\sqrt{-g}K_{{\rm Euler}}{\rm d}^4x$.}.
We calculated the invariant $K_{{\rm Euler}}$ for the Kerr-Newman-(anti-)de Sitter black hole spacetime. The novel explicit algebraic expression of the Euler invariant we computed is:
\begin{align}
K_{{\rm Euler}}&=\frac{1}{3 \left(r^{2}+a^{2} \cos \! \left(\theta \right)^{2}\right)^{6}}\Biggl(-8 \Lambda^{2} \cos \! \left(\theta \right)^{12} a^{12}-48 \Lambda^{2} \cos \! \left(\theta \right)^{10} a^{10} r^{2}\nonumber \\
&-120 \Lambda^{2} \cos \! \left(\theta \right)^{8} a^{8} r^{4}+\left(-160 a^{6} \Lambda^{2} r^{6}+144 a^{6} m^{2}\right) \cos \! \left(\theta \right)^{6}\nonumber \\
&+\left(-120 \Lambda^{2} a^{4} r^{8}-2160 a^{4} m^{2} r^{2}+1440 a^{4} m \,q^{2} r -120 q^{4} a^{4}\right) \cos \! \left(\theta \right)^{4}\nonumber \\
&-2160 \left(\frac{1}{45} \Lambda^{2} a^{2} r^{8}-a^{2} m^{2} r^{2}+\frac{4}{3} a^{2} m \,q^{2} r -\frac{19}{45} a^{2} q^{4}\right) r^{2} \cos \! \left(\theta \right)^{2}\nonumber \\
&-8 r^{12} \Lambda^{2}-144 m^{2} r^{6}+288 m \,q^{2} r^{5}-120 q^{4} r^{4}\Biggr).
\label{LeonardEuler}
\end{align}

We note that the Euler-Poincare characteristic we introduced in Footnote \ref{EulerP}, is also given  as an integral  that involves the Weyl tensor through the invariant $I_1$, and the Ricci invariants $I_5$ and $I_6$ \cite{Avez} \footnote{It is worth mentioning that the topological Euler invariant has been studied in relation to Weyl conformal anomaly in four derivative theories such as conformal gravity and conformal supergravity \cite{duff}  as  well as in the context of boundary conformal invariants in five dimensions \cite{solodukhin}.}:
\begin{align}
-128\pi^2 \chi&=\int (-C_{\alpha\beta\gamma\delta}C^{\alpha\beta\gamma\delta}+2R_{\alpha\beta}R^{\alpha\beta}-\frac{2}{3}R^2)\sqrt{-g}{\rm d}^4 x\\
\label{avezmacaroni}
&\overset{(\ref{dianyanein})}{=}\int (-R_{\alpha\beta\gamma\delta}R^{\alpha\beta\gamma\delta}+4R_{\alpha\beta}R^{\alpha\beta}-R^2)\sqrt{-g}{\rm d}^4 x
\end{align}

 \subsection{Analytic computation of curvature invariants for the Kerr-Newman-(anti-)de Sitter black hole}\label{CIZMKNadSALL}
 In this section we compute the Weyl,Ricci and mixed invariants for the Kerr-Newman-(a)dS black hole. Our analytic results are:
\begin{align}
I_1&=\frac{48}{\left(r^{2}+a^{2}\cos \! \left(\theta \right)^{2}\right)^{6}}\Biggl(-a^{3}m \cos \! \left(\theta \right)^{3}+\left(-3a^{2}m r +a^{2}q^{2}\right)\cos \! \left(\theta \right)^{2}+\left(3a m \,r^{2}-2a \,q^{2}r \right)\cos \! \left(\theta \right)\nonumber \\
&+\left(m r -q^{2}\right)r^{2}\Biggr)\nonumber \\
&\times\left(a^{3}m \cos \! \left(\theta \right)^{3}+\left(-3a^{2}m r +a^{2}q^{2}\right)\cos \! \left(\theta \right)^{2}+\left(-3a m \,r^{2}+2a \,q^{2}r \right)\cos \! \left(\theta \right)-\left(-m r +q^{2}\right)r^{2}\right)\label{CIZMein},\\
I_3&=\frac{96}{\left(r^{2}+a^{2}\cos \! \left(\theta \right)^{2}\right)^{9}}\Biggl(-3\cos \! \left(\theta \right)^{6}a^{6}m^{2}+\left(27a^{4}m^{2}r^{2}-18a^{4}m \,q^{2}r +a^{4}q^{4}\right)\cos \! \left(\theta \right)^{4}\nonumber \\
&+\left(-33a^{2}m^{2}r^{4}+44a^{2}m \,q^{2}r^{3}-14a^{2}q^{4}r^{2}\right)\cos \! \left(\theta \right)^{2}+r^{6}m^{2}-2m \,q^{2}r^{5}+q^{4}r^{4}\Biggr)\nonumber \\
&\times\left(\left(-3a^{2}m r +a^{2}q^{2}\right)\cos \! \left(\theta \right)^{2}+m \,r^{3}-q^{2}r^{2}\right)\\
I_4&=-\frac{864a }{\left(r^{2}+a^{2}\cos \! \left(\theta \right)^{2}\right)^{9}}\left(\frac{\cos \left(\theta \right)^{3}a^{2}m}{3}+\left(-m \,r^{2}+\frac{2}{3}q^{2}r \right)\cos \! \left(\theta \right)\right)\,\nonumber \\
&\times \Biggl[-\frac{\cos \left(\theta \right)^{6}a^{6}m^{2}}{3}+\left(11a^{4}m^{2}r^{2}-\frac{22}{3}a^{4}m \,q^{2}r +a^{4}q^{4}\right)\cos \! \left(\theta \right)^{4}\nonumber \\
&+\left(-9a^{2}m^{2}r^{4}+12a^{2}m \,q^{2}r^{3}-\frac{10}{3}a^{2}q^{4}r^{2}\right)\cos \! \left(\theta \right)^{2}-\frac{r^{4}\left(-3m^{2}r^{2}+6m \,q^{2}r -3q^{4}\right)}{3}\Biggr],\\
I_5&=R=4\Lambda,\\
I_7&=\frac{4\left(\Lambda^{2}\cos \! \left(\theta \right)^{8}a^{8}+4\Lambda^{2}\cos \! \left(\theta \right)^{6}a^{6}r^{2}+6\Lambda^{2}\cos \! \left(\theta \right)^{4}a^{4}r^{4}+4\Lambda^{2}\cos \! \left(\theta \right)^{2}a^{2}r^{6}+\Lambda^{2}r^{8}+3q^{4}\right)\Lambda}{\left(r^{2}+a^{2}\cos \! \left(\theta \right)^{2}\right)^{4}},\label{I7}\\
I_{8}&=\frac{1}{\left(r^{2}+a^{2}\cos \! \left(\theta \right)^{2}\right)^{8}}\Biggl(4\Lambda^{4}\cos \! \left(\theta \right)^{16}a^{16}+32\Lambda^{4}\cos \! \left(\theta \right)^{14}a^{14}r^{2}+112\Lambda^{4}\cos \! \left(\theta \right)^{12}a^{12}r^{4}\nonumber \\
&+224\Lambda^{4}\cos \! \left(\theta \right)^{10}a^{10}r^{6}+\left(280\Lambda^{4}a^{8}r^{8}+24\Lambda^{2}a^{8}q^{4}\right)\cos \! \left(\theta \right)^{8}+96r^{2}\left(\frac{7}{3}a^{6}\Lambda^{4}r^{8}+a^{6}q^{4}\Lambda^{2}\right)\cos \! \left(\theta \right)^{6}\nonumber \\
&+\left(112\Lambda^{4}a^{4}r^{12}+144\Lambda^{2}a^{4}q^{4}r^{4}\right)\cos \! \left(\theta \right)^{4}+96r^{2}\left(\frac{1}{3}\Lambda^{4}a^{2}r^{12}+\Lambda^{2}a^{2}q^{4}r^{4}\right)\cos \! \left(\theta \right)^{2}\nonumber \\
&+4\Lambda^{4}r^{16}+24\Lambda^{2}q^{4}r^{8}+4q^{8}\Biggr)\label{vierriccikrummung},
\end{align}

\begin{align}
I_6&=\frac{4q^4}{\left(r^2+a^2 \cos \! \left(\theta \right)^{2}\right)^{4}}+4\Lambda^2,\\
I_9&=-\frac{16\left(-3a^{2}\cos \! \left(\theta \right)^{2}m r +a^{2}\cos \! \left(\theta \right)^{2}q^{2}+m \,r^{3}-q^{2}r^{2}\right)q^{4}}{\left(r^{2}+a^{2}\cos \! \left(\theta \right)^{2}\right)^{7}},\\
I_{10}&=\frac{16 a q^{4} }{\left(r^{2}+a^{2}\cos \! \left(\theta \right)^{2}\right)^{7}}\left(\cos \! \left(\theta \right)^{3}a^{2}m -3\cos \! \left(\theta \right)m \,r^{2}+2\cos \! \left(\theta \right)q^{2}r \right),\\
I_{11}&=\frac{64q^{4}}{\left(r^{2}+a^{2}\cos \! \left(\theta \right)^{2}\right)^{10}}\Biggl[\cos \! \left(\theta \right)^{3}a^{3}m +\left(-3a^{2}m r +a^{2}q^{2}\right)\cos \! \left(\theta \right)^{2}+\left(-3a m \,r^{2}+2a \,q^{2}r \right)\cos \! \left(\theta \right)\nonumber \\
&-r^{2}\left(-m r +q^{2}\right)\Biggr]\nonumber \\
&\times \left(-\cos \! \left(\theta \right)^{3}a^{3}m +\left(-3a^{2}m r +a^{2}q^{2}\right)\cos \! \left(\theta \right)^{2}+\left(3a m \,r^{2}-2a \,q^{2}r \right)\cos \! \left(\theta \right)+\left(m r -q^{2}\right)r^{2}\right),\\
I_{12}&=-\frac{128 a q^{4}}{\left(r^{2}+a^{2}\cos \! \left(\theta \right)^{2}\right)^{10}} \left(\cos \! \left(\theta \right)^{3}a^{2}m -3\cos \! \left(\theta \right)m \,r^{2}+2\cos \! \left(\theta \right)q^{2}r \right)\nonumber \\
&\times \left(-3a^{2}\cos \! \left(\theta \right)^{2}m r +a^{2}\cos \! \left(\theta \right)^{2}q^{2}+m \,r^{3}-q^{2}r^{2}\right),\\
I_{15}&=\frac{4q^{4}\left(\cos \! \left(\theta \right)^{2}a^{2}m^{2}+r^{2}m^{2}-2m \,q^{2}r +q^{4}\right)}{\left(r^{2}+a^{2}\cos \! \left(\theta \right)^{2}\right)^{8}},\\
I_{16}&=-\frac{24q^{4}\left(\cos \! \left(\theta \right)^{2}a^{2}m^{2}+\left(m r -q^{2}\right)^{2}\right)\left(a^{2}\left(m r -\frac{q^{2}}{3}\right)\cos \! \left(\theta \right)^{2}-\frac{r^{2}\left(m r -q^{2}\right)}{3}\right)}{\left(r^{2}+a^{2}\cos \! \left(\theta \right)^{2}\right)^{11}},\\
I_{17}&=-\frac{24 a\,q^{4}}{\left(r^{2}+a^{2}\cos \! \left(\theta \right)^{2}\right)^{11}}\,\left(\frac{\cos \left(\theta \right)^{3}a^{2}m}{3}+\left(-m \,r^{2}+\frac{2}{3}q^{2}r \right)\cos \! \left(\theta \right)\right) \,\left(\cos \! \left(\theta \right)^{2}a^{2}m^{2}+(mr-q^2)^2\right).
\label{CIsiebenundzehn}
\end{align}

We summarise our results as follows:
\begin{theorem}\label{courbureinvriemein}
The exact algebraic expressions for the curvature invariants calculated for the Kerr-Newman-(anti-)de Sitter metric are given in Equations (\ref{CIZMein})-(\ref{CIsiebenundzehn}) and (\ref{CHERNPOTRYAGINNOBESCH}).
\end{theorem}

\begin{corollary}
For zero cosmological constant ($\Lambda=0$), our results reduce correctly to those obtained in \cite{overduincwh} for the Kerr-Newman metric.
\end{corollary}

\begin{figure}[ptbh]
\centering
  \begin{subfigure}[b]{.60\linewidth}
    \centering
    \includegraphics[width=.99\textwidth]{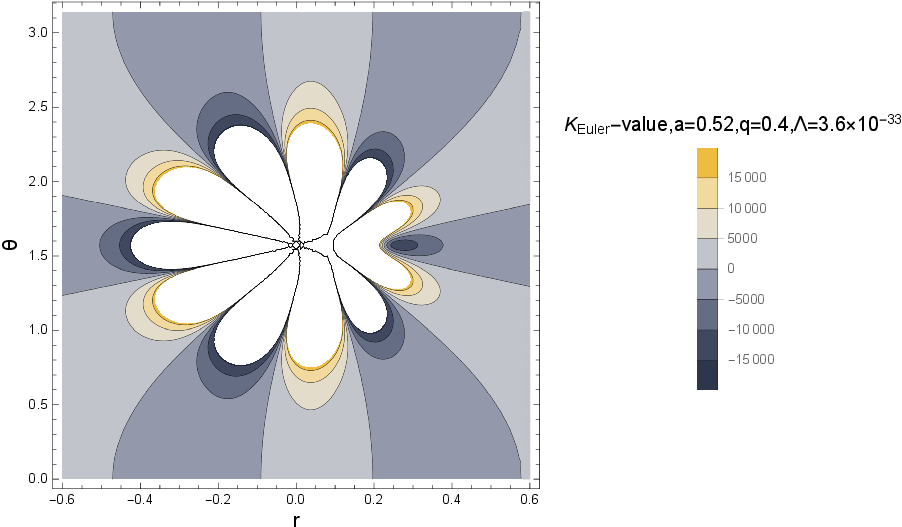}
    \caption{ Contour plot of  $K_{{\rm Euler}}$.}\label{ContourEulera052q04L011}
  \end{subfigure}%
  \begin{subfigure}[b]{.60\linewidth}
    \centering
    \includegraphics[width=.99\textwidth]{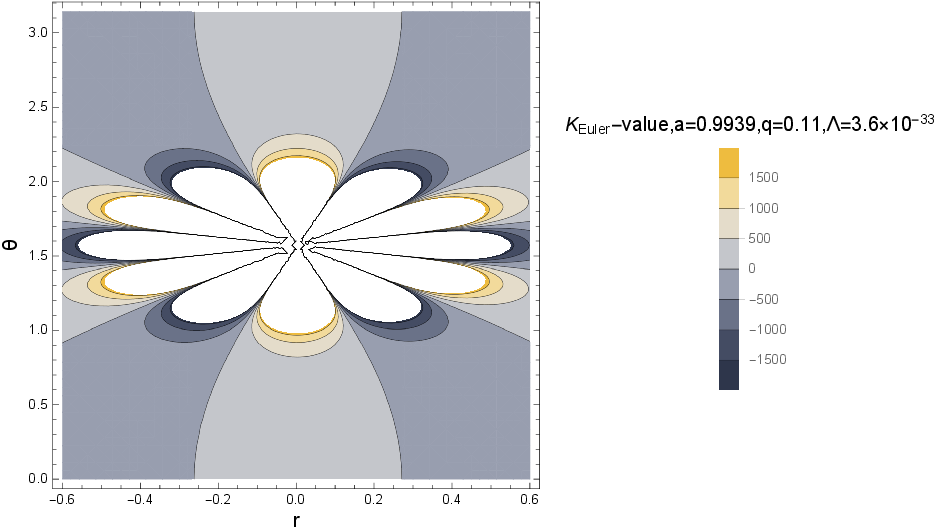}
    \caption{Contour plot of $K_{{\rm Euler}}$}\label{ContourEulera09939q011L011}
  \end{subfigure}\\
  \begin{subfigure}[b]{.60\linewidth}
    \centering
    \includegraphics[width=.99\textwidth]{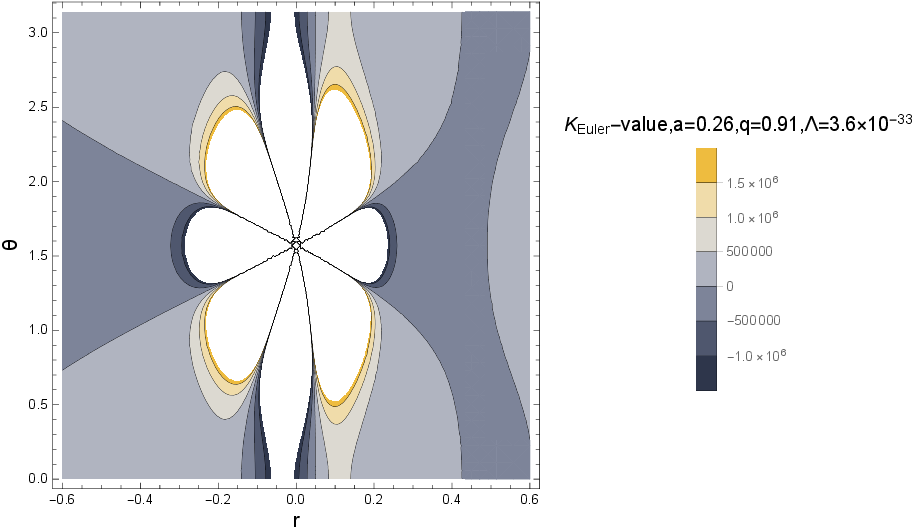}
    \caption{Contour Plot of $K_{{\rm Euler}}$.}\label{ContourEulera026q091L011}
  \end{subfigure}%
  \caption{Contour plots of  $K_{{\rm Euler}}$. (a) for spin parameter $a=0.52$, charge $q=0.4$,dimensionless cosmological parameter $\Lambda=3.6\times 10^{-33},m=1$. (b) For spin  $a=0.9939$, charge $q=0.11$,dimensionless cosmological parameter $\Lambda=3.6\times 10^{-33},m=1$. (c) For low spin  $a=0.26$, electric charge $q=0.91$,dimensionless cosmological parameter $\Lambda=3.6\times 10^{-33}$ and mass $m=1$.}\label{ContourPlotsEulerLeonard}
\end{figure}

\begin{figure}[ptbh]
\centering
  \begin{subfigure}[b]{.60\linewidth}
    \centering
    \includegraphics[width=.99\textwidth]{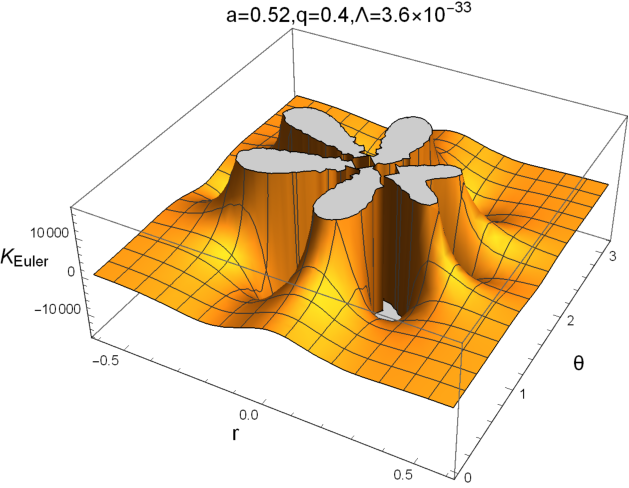}
    \caption{ 3D plot of  $K_{{\rm Euler}}$.}\label{3dKEulera052q04L001}
  \end{subfigure}%
  \begin{subfigure}[b]{.60\linewidth}
    \centering
    \includegraphics[width=.99\textwidth]{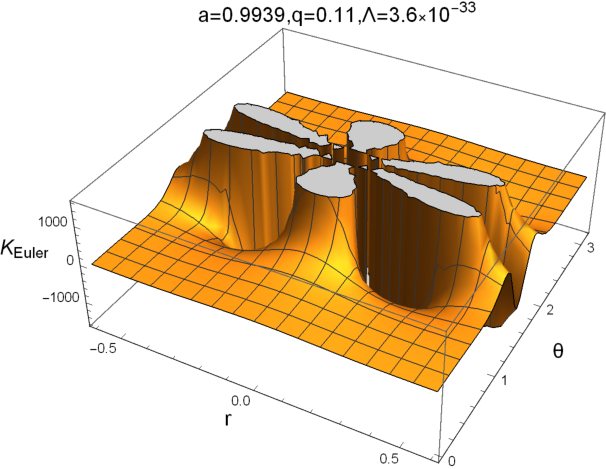}
    \caption{3D plot of $K_{{\rm Euler}}$}\label{3dKEulera09939q011L001}
  \end{subfigure}\\
  \begin{subfigure}[b]{.60\linewidth}
    \centering
    \includegraphics[width=.99\textwidth]{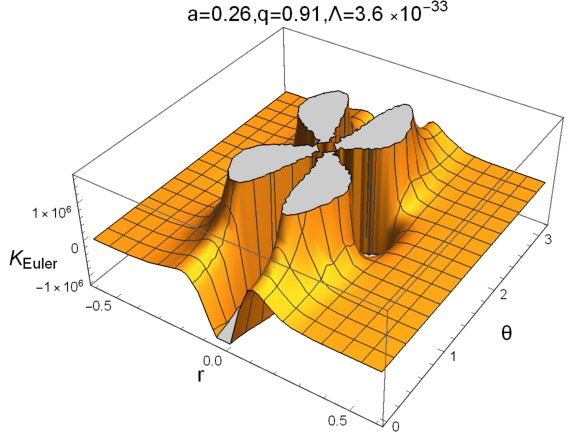}
    \caption{3D Plot of $K_{{\rm Euler}}$.}\label{3dKEulera026q091L001}
  \end{subfigure}%
  \caption{3D plots of the Euler invariant,  $K_{{\rm Euler}}$, plotted as a function of the Boyer-Lindquist coordinates $r$ and $\theta$ . (a) for spin parameter $a=0.52$, charge $q=0.4$,dimensionless cosmological parameter $\Lambda=3.6\times 10^{-33},m=1$. (b) For spin  $a=0.9939$, charge $q=0.11$,dimensionless cosmological parameter $\Lambda=3.6\times 10^{-33},m=1$. (c) For low spin  $a=0.26$, electric charge $q=0.91$,dimensionless cosmological parameter $\Lambda=3.6\times 10^{-33}$ and mass $m=1$.}\label{TRIADIMKEuler}
\end{figure}

In Figure \ref{ContourPlotsEulerLeonard} we plot level curves of the Euler invariant  $K_{{\rm Euler}}$, in the $r-\theta$ space, for different sets of values of the physical black hole parameters $a,q,\Lambda,m$ \footnote{With regard to the spin $a$ (Kerr parameter) we note that observations of near-infrared periodic flares have revealed that the central black
hole $SgrA^{*}$ is rotating with a reported spin parameter: $a=0.52 (\pm 0.1,\pm 0.08,\pm 0,08)$ \cite{genzelr}. The error estimates here the uncertainties in the
period, black hole mass and distance to the galactic centre, respectively. Observation of X-ray flares confirmed that the spin of the supermassive black hole is
indeed substantial and values of the Kerr parameter as high as: $a=0.9939^{+0.0026}_{-0.0074}$ have been obtained \cite{aschenbach}. For our plots we choose values for the Kerr parameter consistent with these observations.}.
In Figure \ref{TRIADIMKEuler} we display three-dimensional plots of the curvature invariant $K_{{\rm Euler}}$ as a function of the Boyer-Lindquist coordinates $r$ and $\theta$, for three sets of values for the spin, electric charge, cosmological constant and mass of the black hole. It is evident that the geometry of spacetime, as exhibited by the plotting of $K_{{\rm Euler}}$, inside the KN(a)dS black hole is far from simple. We also observe a broader variation along the $r-$direction than in the $\theta-$direction for moderate and high spin values, Figures (\ref{3dKEulera052q04L001})-(\ref{3dKEulera09939q011L001}), as opposed to a narrower variation along the $r-$direction and broader variation along the $\theta-$direction for
small spin values and high values of the electric charge- see Figure (\ref{3dKEulera026q091L001}).

\begin{figure}[ptbh]
\centering
  \begin{subfigure}[b]{.60\linewidth}
    \centering
    \includegraphics[width=.99\textwidth]{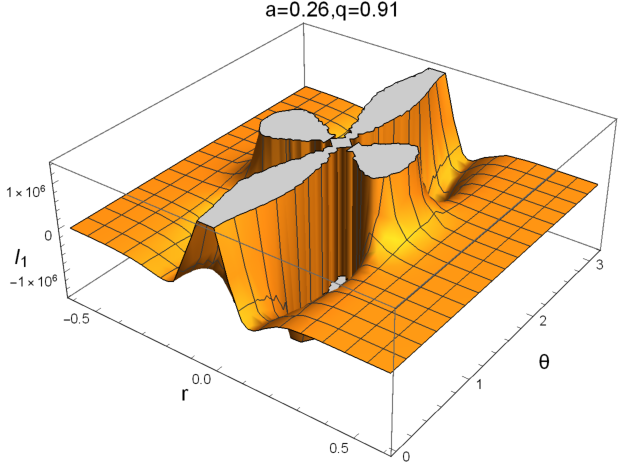}
    \caption{ 3D plot of  $I_1$.}\label{graphI1a06q08}
  \end{subfigure}%
  \begin{subfigure}[b]{.60\linewidth}
    \centering
    \includegraphics[width=.99\textwidth]{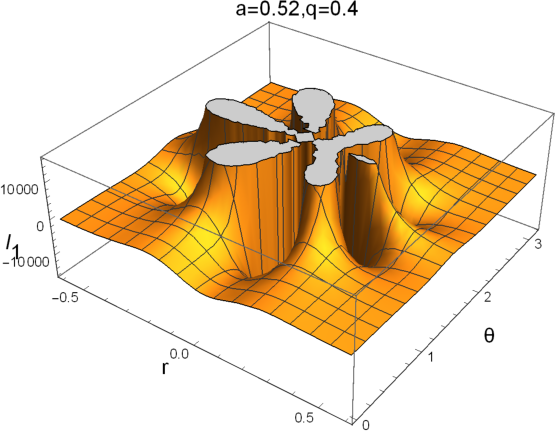}
    \caption{3D plot of $I_1$}\label{graphWI1a052q04}
  \end{subfigure}\\
  \begin{subfigure}[b]{.60\linewidth}
    \centering
    \includegraphics[width=.99\textwidth]{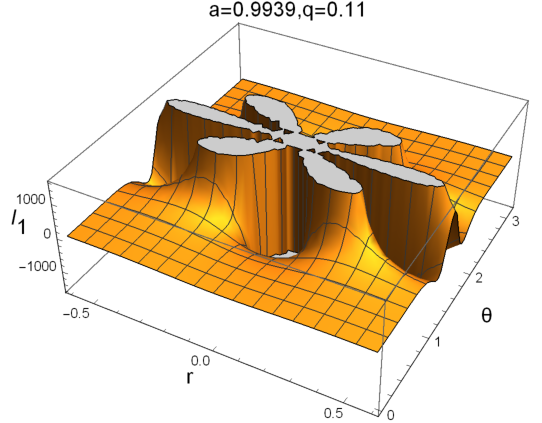}
    \caption{3D Plot of $I_1$.}\label{grafosI1W09939011}
  \end{subfigure}%
  \caption{3D plots of the ZM Weyl invariant,  $I_1$, plotted as a function of the Boyer-Lindquist coordinates $r$ and $\theta$ . (a) for spin parameter $a=0.26$, charge $q=0.91$, $m=1$. (b) For spin  $a=0.52$, charge $q=0.4$, $m=1$. (c) For high spin  $a=0.9939$, electric charge $q=0.11$  and mass $m=1$.}\label{TRIADI1Weyl}
\end{figure}

\begin{figure}[ptbh]
\centering
  \begin{subfigure}[b]{.60\linewidth}
    \centering
    \includegraphics[width=.99\textwidth]{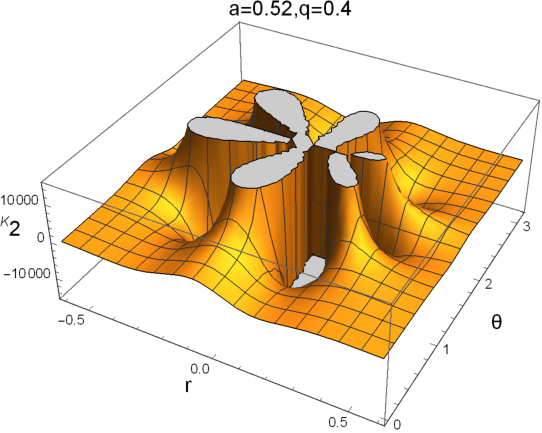}
    \caption{ 3D plot of  $K_2$.}\label{grafosHirzebruch052q04}
  \end{subfigure}%
   \begin{subfigure}[b]{.60\linewidth}
    \centering
    \includegraphics[width=.99\textwidth]{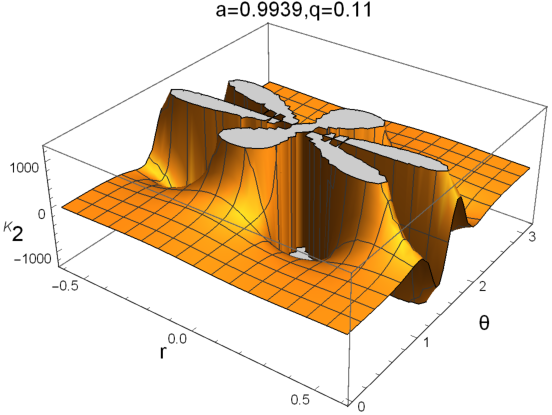}
    \caption{3D Plot of $K_2$.}\label{diagrCPH09939011}
  \end{subfigure}%
  \caption{3D plots of the Chern-Pontryagin invariant,  $K_2$, plotted as a function of the Boyer-Lindquist coordinates $r$ and $\theta$ . (a) for spin parameter $a=0.52$, charge $q=0.4$, mass $m=1$. (b) For spin  $a=0.9939$, charge $q=0.11$, mass $m=1$. }\label{TRIADK2CP}
\end{figure}

In Figures \ref{graphI1a06q08}-\ref{diagrCPH09939011} we plot the first two invariants for various values of the Kerr parameter and electric charge of the black hole with respect to $r,\theta$ coordinates.

\begin{figure}[ptbh]
\centering
  \begin{subfigure}[b]{.60\linewidth}
    \centering
    \includegraphics[width=.99\textwidth]{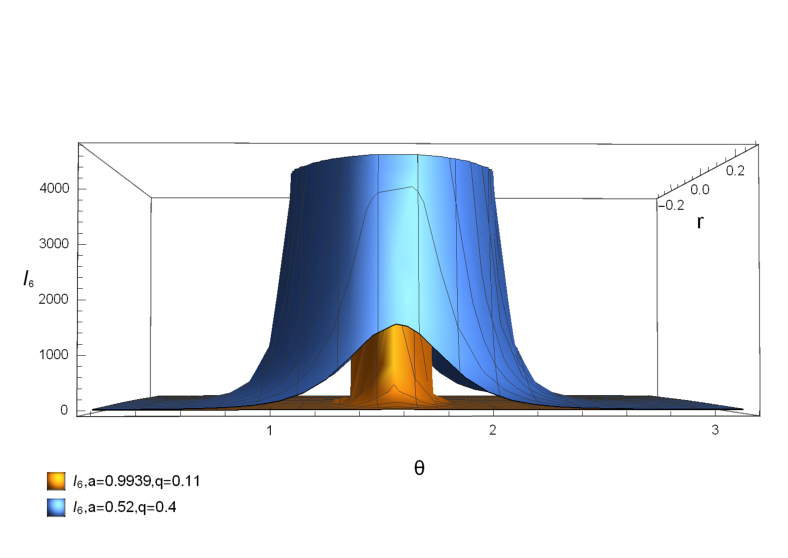}
    \caption{ 3D plot of  $I_6$.}\label{diagrI6Ricci}
  \end{subfigure}%
   \begin{subfigure}[b]{.60\linewidth}
    \centering
    \includegraphics[width=.99\textwidth]{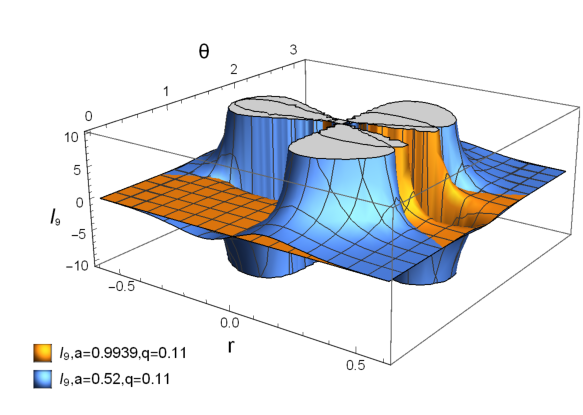}
    \caption{3D Plot of $I_9$.}\label{diagrI9a09939052q011}
  \end{subfigure}\\
  \begin{subfigure}[b]{.60\linewidth}
    \centering
    \includegraphics[width=.99\textwidth]{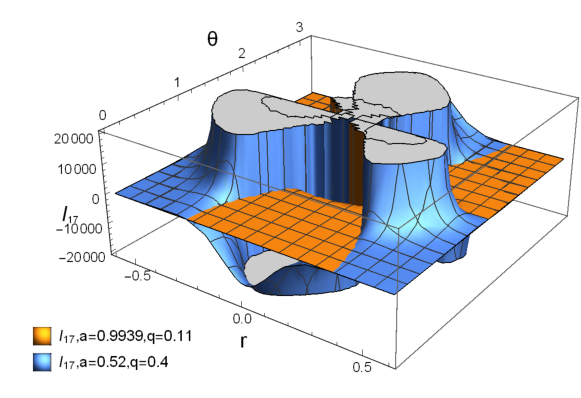}
    \caption{3D Plot of $I_{17}$.}\label{diagrI17MixedWeyl}
  \end{subfigure}
  \caption{The invariants $I_6,I_9,I_{17}$ all plotted  as a function of the Boyer-Lindquist coordinates $r$ and $\theta$. (a)The Ricci invariant $I_6$ plotted as a function  of the coordinates $r$ and $\theta$ for a black hole of mass $m=1$, for two different pairs of values for Kerr parameter and charge $(a=0.9939,q=0.11),(a=0.52,q=0.4)$ and fixed value for the cosmological constant $\Lambda=3.6\times 10^{-33}$.(b) The mixed invariant $I_9$ is plotted as a function of the coordinates $r$ and $\theta$ for a black hole of mass $m=1$, for two different values of the Kerr parameter $a=0.9939,a=0.52$ and electric charge $q=0.11$.(c) The mixed invariant $I_{17}$ plotted as a function of the Boyer-Lindquist coordinates $r$ and $\theta$ for a black hole of mass $m=1$, for two different pairs of values for the Kerr parameter and electric charge: $(a=0.9939,q=0.11),(a=0.52,q=0.4)$. }\label{TRIADI6I9I17}
\end{figure}

In Figure \ref{diagrI6Ricci} we plot the Ricci invariant $I_6$  as a function of the Boyer-Lindquist coordinates $r$ and $\theta$ for a black hole of mass $m=1$, for two different pairs of values for Kerr parameter and electric charge: $(a=0.9939,q=0.11),(a=0.52,q=0.4)$ and a fixed value for a dimensionless positive cosmological constant. In the context of setting $m=1$, a dimensionless form of $\Lambda$ corresponds to the dimensionless combination $m^2\Lambda$, unless otherwise stipulated. This means that for supermassive black holes such as at the centre of Galaxy M87 with mass $M^{M87}_{\rm BH}=6.7\times 10^9$ solar masses \cite{EHT} the value of dimensionless $\Lambda=3\times 10^{-4}$ corresponds to the value for the cosmological constant: $\Lambda=3.06\times 10^{-34}{\rm cm}^{-2}$. For the galactic centre supermassive SgrA* black hole with mass $M_{\rm BH}^{SgrA*}=4.06\times 10^6$ solar masses \cite{Eisenhauer} a value for dimensionless $\Lambda=3.6\times 10^{-33}$ (which is the value we use in our graphs) corresponds to the value for cosmological constant $10^{-56}{\rm cm}^{-2}$ consistent with observations.

\begin{figure}
[ptbh]
\begin{center}
\includegraphics[height=2.4526in, width=3.3797in ]{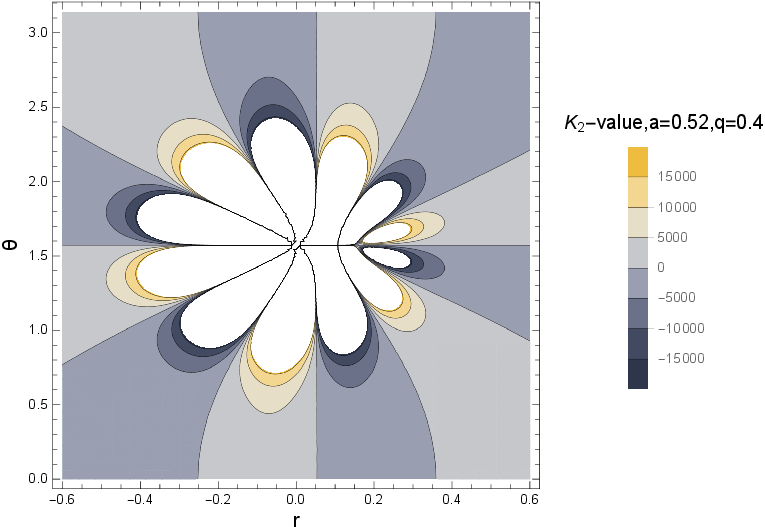}
 \caption{Contour plot of level curves for the Chern-Pontryagin invariant $K_2$ for $a=0.52,q=0.4,m=1$. }%
\label{levelK2a052q04}%
\end{center}
\end{figure}

In Figures \ref{diagrI9a09939052q011}, \ref{diagrI17MixedWeyl}, we plot the mixed curvature invariants $I_9$ and $I_{17}$ respectively, as a function of the Boyer-Lindquist coordinates $r$ and $\theta$.
In Figure \ref{levelK2a052q04} we plot the level sets (isoline) of the topological Chern-Pontryagin curvature invariant $K_2$.

As it is evident from the plots in Figures \ref{ContourPlotsEulerLeonard}-\ref{diagrI17MixedWeyl},\ref{levelK2a052q04} the geometry of spacetime inside Kerr-Newman-(anti-)de Sitter black holes exhibit a rich structure. We also observe that curvature is negative over significant regions of the $r-\theta$ space, a result that is consistent with earlier findings for simpler spacetimes obtained by several researchers  \cite{schmidt}, \cite{LakeK}, while is particularly extreme near the ring singularity at $r=0,\theta=\pi/2$. The appearance of fluctuations in the Kerr spacetime have been attributed to conflicting contributions to the curvature from the  gravitoelectric and gravitomagnetic components of the Weyl tensor,  the latter components generated by the black hole's rotation \cite{Filipe},\cite{cbcruf}.
In section \ref{eswterikomelanisopis} we will determine the sign of the ZM curvature invariants, i.e. we shall demarcate the regions in the $r-\theta$ space of positive and negative values for the corresponding invariants for the more general case of the accelerating Kerr-Newman black hole in (anti-)de Sitter spacetime (figures 10-16 and figure 20).

\subsection{Curvature invariants and the Newman-Penrose formalism}\label{RogerEwa}

The Newman-Penrose (NP) formalism is a tetrad formalism with a special choice of the tetrad in terms the null vectors $\mathbf{l},\mathbf{n},\mathbf{m},\mathbf{\overline{m}}$ \cite{NPformalismCI}.
In the NP formalism, the ten independent components of the Weyl tensor are determined by the five complex scalar functions defined as:
\begin{align}
\Psi_0&=C_{\mu\nu\lambda\sigma}l^{\mu}m^{\nu}l^{\lambda}m^{\sigma},\nonumber \\
\Psi_1&=C_{\mu\nu\lambda\sigma}l^{\mu}n^{\nu}l^{\lambda}m^{\sigma},\nonumber \\
\Psi_2&=C_{\mu\nu\lambda\sigma}\overline{m}^{\mu}n^{\nu}l^{\lambda}m^{\sigma},\\
\Psi_3&=C_{\mu\nu\lambda\sigma}\overline{m}^{\mu}n^{\nu}l^{\lambda}n^{\sigma},\nonumber \\
\Psi_4&=C_{\mu\nu\lambda\sigma}\overline{m}^{\mu}n^{\nu}\overline{m}^{\lambda}n^{\sigma}.\nonumber
\end{align}

Two particularly useful complex scalar polynomial invariants for a vacuum spacetime are given in terms of the Weyl tensor components by \cite{GrifPod}:
\begin{equation}
\mathbf{I}=\Psi_0\Psi_4-4\Psi_1\Psi_3+3\Psi_2^2,\;\;\mathbf{J}=\det\Psi,\;\;\Psi=\left[\begin{array}{ccc}
\Psi_0 & \Psi_1 & \Psi_2 \\
\Psi_1 & \Psi_2 & \Psi_3 \\
\Psi_2 & \Psi_3 & \Psi_4\end{array}\right].
\end{equation}
As a matter of fact, as we will show in what follows, the real part of $\mathbf{I}$ is $\frac{1}{16}$ times the Kretschmann scalar for the vacuum case. For the non-vacuum spacetimes which is the class of black hole spacetimes we investigate in this work we will see how the Chern-Pontryagin invariant $K_2$ and Weyl invariant $I_1$ are related to $\mathbf{I}$.

For our computations we use the generalised Kinnersley null tetrad used in \cite{KraniotisDirac}, for the construction of the generalised massive Dirac equation in the Kerr-Newman-(anti-)de Sitter black hole background:

\begin{align}
l^{\mu}&=\left[\frac{(r^2+a^2)\Xi}{\Delta_r^{KN}},1,0,\frac{a\Xi}{\Delta_r^{KN}}\right],\;
n^{\mu}=\left[\frac{\Xi(r^2+a^2)}{2\rho^2},-\frac{\Delta_r^{KN}}{2\rho^2},0,\frac{a\Xi}{2\rho^2}\right] \nonumber \\
m^{\mu}&=\frac{1}{(r+ia\cos\theta)\sqrt{2\Delta_{\theta}}}\left[ia\Xi\sin\theta,0,\Delta_{\theta},\frac{i\Xi}{\sin\theta}\right]\nonumber \\
\overline{m}^{\mu}&=\frac{-1}{(r-ia\cos\theta)\sqrt{2\Delta_{\theta}}}\left[ia\Xi\sin\theta,0,-\Delta_{\theta},\frac{i\Xi}{\sin\theta}\right]
\label{GLAMBDAKNNULLTETRAD}
\end{align}

From (1.3a) in \cite{CarminatiMcL} we have the relation for the trace-free Ricci tensor $S_{\alpha\beta}$, in terms of the Ricci scalars:
\begin{align}
\frac{1}{4}S_{\alpha}^{\;\beta}S_{\beta}^{\;\alpha}=2[\Phi_{20}\Phi_{02}+\Phi_{22}\Phi_{00}-2\Phi_{12}\Phi_{10}-2
\Phi_{21}\Phi_{01}+2\Phi_{11}^2,]
\end{align}
and
\begin{align}
R_{\alpha}^{\;\beta}R_{\beta}^{\;\alpha}&=S_{\alpha}^{\;\beta}S_{\beta}^{\;\alpha}+6.24\Lambda^{\prime 2},\nonumber\\
&=S_{\alpha}^{\;\beta}S_{\beta}^{\;\alpha}+8.18\Lambda^{\prime 2},
\end{align}
where $R=24\Lambda^{\prime}$ and $\Lambda^{\prime}$ is the Ricci scalar introduced in \cite{Chandrasekhar}.
Thus, we obtain:
\begin{align}
I_6\equiv R_{\alpha}^{\;\beta}R_{\beta}^{\;\alpha}&=8\Biggl[\Phi_{20}\Phi_{02}+\Phi_{22}\Phi_{00}
-2 \Phi_{12}\Phi_{10}-2\Phi_{21}\Phi_{01}\nonumber \\
&+2 \Phi_{11}^2+18\Lambda^{\prime 2}\Biggr]
\label{scalarsRicci}
\end{align}

Using the generalised Kinnersley null tetrad we compute the Ricci scalars in Eqn.(\ref{scalarsRicci}) with the result:
\begin{align}
R_{\alpha}^{\;\beta}R_{\beta}^{\;\alpha}&=16\Phi_{11}^2+8.18\Lambda^{\prime2}\nonumber \\
&=16\left(\frac{q^2}{2(r^2+a^2\cos(\theta)^2)^2}\right)^2+8.18\left(\frac{1}{24}4\Lambda\right)^2\nonumber \\
&=\frac{4q^4}{\left(r^2+a^2\cos(\theta)^2\right)^4}+4\Lambda^2,
\end{align}
where we computed for the Ricci scalars:
\begin{align}
\Phi_{00}&\equiv\frac{1}{2}R_{\mu\nu}l^{\mu}l^{\nu}=0,\;\Phi_{01}\equiv\frac{1}{2}l^{\mu}m^{\nu}=\overline{\Phi}_{10}=0,\label{glricciein}\\
\Phi_{02}&\equiv\frac{1}{2}R_{\mu\nu}m^{\mu}m^{\nu}=\overline{\Phi}_{20}=0,\;\Phi_{22}\equiv\frac{1}{2}R_{\mu\nu}n^{\mu}n^{\nu}=0,\\
\Phi_{12}&\equiv\frac{1}{2}R_{\mu\nu}n^{\mu}m^{\nu}=\overline{\Phi}_{21}=0,\\
\Phi_{11}&\equiv\frac{1}{4}R_{\mu\nu}(l^{\mu}n^{\nu}+m^{\mu}\overline{m}^{\nu})=\frac{q^2}{2(r^2+a^2\cos(\theta)^2)^2}
\label{glriccifunf}
\end{align}

Now
\begin{align}
S_{\mu}^{\;\nu}S_{\nu}^{\;\rho}S_{\rho}^{\;\mu}&=48\Biggl(-\Phi_{00}\Phi_{11}\Phi_{22}+\Phi_{00}\Phi_{12}\Phi_{21}+
\Phi_{11}\Phi_{02}\Phi_{20}+\Phi_{22}\Phi_{01}\Phi_{10} \nonumber \\
&-\Phi_{01}\Phi_{20}\Phi_{12}-\Phi_{10}\Phi_{02}\Phi_{21}\Biggr),
\end{align}
and
\begin{align}
S_{\mu}^{\;\nu}S_{\nu}^{\;\rho}S_{\rho}^{\;\mu}=R_{\mu}^{\;\nu}R_{\nu}^{\;\rho}R_{\rho}^{\;\mu}
-\frac{3}{4}R R_{\mu}^{\;\nu}R_{\nu}^{\;\mu}+\frac{3R^3}{16}-\frac{4R^3}{64}.
\end{align}
Thus we obtain for the invariant $I_7$ in the NP-formalism the result:
\begin{align}
R_{\mu}^{\;\nu}R_{\nu}^{\;\rho}R_{\rho}^{\;\mu}&=\frac{3}{4}(4\Lambda)\{\frac{4q^4}{\left(r^2+a^2 \cos \! \left(\theta \right)^{2}\right)^{4}}+4\Lambda^2\}-\frac{8}{64}(4\Lambda)^3\nonumber \\
&=4\Lambda^3+\frac{4\Lambda\; 3 q^4}{\left(r^2+a^2 \cos \! \left(\theta \right)^{2}\right)^{4}}.
\label{riccitriploun}
\end{align}
This agrees with our analytic direct computation for $I_7$ in eqn (\ref{I7} )performed with Maple. In producing eqn.(\ref{riccitriploun}) we used the fact that: $S_{\mu}^{\;\nu}S_{\nu}^{\;\rho}S_{\rho}^{\;\mu}\overset{(\ref{glricciein})-(\ref{glriccifunf})}{=}0,$ and $R=4\Lambda$.

Now let us compute the invariant $I_8$ in the NP formalism. We first derive the following relation:
\begin{align}
S_{\mu}^{\;\nu}S_{\nu}^{\;\rho}S_{\rho}^{\lambda}S_{\lambda}^{\mu}&=
R_{\mu}^{\;\nu}R_{\nu}^{\;\rho}R_{\rho}^{\lambda}R_{\lambda}^{\mu}-R_{\mu}^{\;\nu}R_{\nu}^{\;\rho}R_{\rho}^{\;\mu}R+
\frac{3}{8}R_{\mu}^{\;\nu}R_{\nu}^{\;\mu}R^2-\frac{3}{64}R^4\nonumber \\
&=I_8-I_7I_5+\frac{3}{8}I_6 I_5^2-\frac{3}{64}I_5^4.
\label{endiamei8np}
\end{align}
Using our computations in (\ref{glricciein})-(\ref{glriccifunf}) we obtain:
\begin{equation}
S_{\mu}^{\;\nu}S_{\nu}^{\;\rho}S_{\rho}^{\lambda}S_{\lambda}^{\mu}=64\Phi_{11}^4=64\left[\frac{q^2}{2(r^2+a^2\cos(\theta)^2)^2}\right]^4.
\end{equation}
Thus, solving eqn(\ref{endiamei8np}) for the curvature invariant $I_8$ yields:
\begin{align}
I_8&=\frac{4 q^{8}}{\left(r^{2}+a^{2} \cos \! \left(\theta \right)^{2}\right)^{8}}+4 \left(4 \Lambda^{3}+\frac{12 \Lambda  q^{4}}{\left(r^{2}+a^{2} \cos \! \left(\theta \right)^{2}\right)^{4}}\right) \Lambda -6 \left(\frac{4 q^{4}}{\left(r^{2}+a^{2} \cos \! \left(\theta \right)^{2}\right)^{4}}+4 \Lambda^{2}\right) \Lambda^{2}\nonumber\\
&+12 \Lambda^{4},
\end{align}
a result in total agreement with our brute force tensorial computations in eqn.(\ref{vierriccikrummung}).

The Weyl scalar $\Psi_2=C_{\mu\nu\lambda\sigma}\overline{m}^{\mu}n^{\nu}l^{\lambda}m^{\sigma}$ is calculated to be:
\begin{equation}
\Psi_2=-\frac{i \cos(\theta)ma+mr-q^2}{(r-ia\cos(\theta))^3(r+i a\cos(\theta))}.
\label{zweipsi}
\end{equation}
We note that Eqn(\ref{zweipsi}) for zero electric charge ($q=0$) reduces correctly to the expression derived for Kerr spacetime in \cite{Chandrasekhar}:
\begin{equation}
\Psi_2=-\frac{m}{(\overline{\rho}^*)^3}, \;\;\;\overline{\rho}^*=r-ia\cos(\theta).
\end{equation}

A non-trivial check of our analytic computations performed in section, provided with the aid of the NP formalism, is the following equation that relates the Weyl invariant $I_1$ and the Chern-Potryagin invariant $K_2$ with the Weyl scalar $\Psi_2$ computed in eqn(\ref{zweipsi}):
\begin{equation}
\mathbb{I}:=I_1-iK_2=16.3\Psi_2^2=16\mathbf{I}.
\end{equation}
We also derived the following relation:
\begin{equation}
\mathbb{J}:=I_3+iI_4=96(-\Psi_2)^3=6.16(-\Psi_2)^3=96\mathbf{J}.
\end{equation}

A more detailed account of the use of the NP formalism for the computation of all curvature invariants for rotating charged black holes with $\Lambda\not =0$ will be a subject of a separate publication.

\subsubsection{The syzygies}

As the \textit{complete} set of Riemann invariants in \cite{CarminatiMcL} contains more than 14 elements not all of them are independent for the different Petrov and Segre types of spacetimes. The relations between the dependent and independent invariants for the different types are called \textit{syzygies} \cite{spinors}.  Such relations provide further non-trivial consistency checks for our analytic computations.
The first non-trivial such type of relation we derive for the Kerr-Newman-(anti-)de Sitter black hole is:
\begin{equation}
\mathbb{I}^3=12\mathbb{J}^2.
\end{equation}
The second highly non-trivial syzygy for the Ricci invariants we derive in this work for the case of the Kerr-Newman black hole in the presence of the cosmological constant $\Lambda$ is:
\begin{equation}
I_6^2=4I_8-\frac{4}{3}I_7I_5+\frac{1}{12}I_5^4.
\label{syzyg1}
\end{equation}
Defining the quantities:
\begin{align}
\mathbb{L}=I_{11}+iI_{12},\;\;\mathbb{K}=I_9+iI_{10},
\end{align}
we derive the syzygy:
\begin{equation}
3\mathbb{L}^2=\mathbb{I}\;\mathbb{K}^2.
\label{syzygzwei}
\end{equation}
We also discovered the following syzygies:
\begin{align}
16(I_6-\frac{1}{4}I_5^2)\mathbb{M}_1&=\mathbb{K}\overline{\mathbb{K}}=|\mathbb{K}|^2,\label{syzdrei}\\
3072(I_6-\frac{1}{4}I_5^2)^2\mathbb{M}_2^2&=\mathbb{I}\mathbb{K}^2\overline{\mathbb{K}}^2\label{syzvier},
\end{align}
where we define:
\begin{equation}
\mathbb{M}_1:=I_{15},\;\;\mathbb{M}_2:=I_{16}+iI_{17}.
\end{equation}
For completeness we provide the following analytic expressions for the quantities appearing in the syzygies:
\begin{align}
&\mathbb{I}=-\frac{96\left(\frac{\cos \left(\theta \right)^{2}a^{2}m^{2}}{2}+\left(-\mathrm{i}a \,m^{2}r +\mathrm{i}m \,q^{2}a \right)\cos \! \left(\theta \right)-\frac{m^{2}r^{2}}{2}+m \,q^{2}r -\frac{q^{4}}{2}\right)}{\left(r^{2}+a^{2}\cos \! \left(\theta \right)^{2}\right)^{2}\left(a^{4}\cos \! \left(\theta \right)^{4}+4\,\mathrm{i}\cos \! \left(\theta \right)^{3}a^{3}r -6\cos \! \left(\theta \right)^{2}a^{2}r^{2}-4\,\mathrm{i}\cos \! \left(\theta \right)a \,r^{3}+r^{4}\right)} \nonumber \\
&=-\frac{96\left(\frac{\cos \left(\theta \right)^{2}a^{2}m^{2}}{2}+\left(-\mathrm{i}a \,m^{2}r +\mathrm{i}m \,q^{2}a \right)\cos \! \left(\theta \right)-\frac{m^{2}r^{2}}{2}+m \,q^{2}r -\frac{q^{4}}{2}\right)}{\left(r^{2}+a^{2}\cos \! \left(\theta \right)^{2}\right)^{2}\left(r-ia \cos(\theta)\right)^4},\\
&\mathbb{L}=-\frac{128\left(\frac{\cos \left(\theta \right)^{2}a^{2}m^{2}}{2}+\left(-\mathrm{i}a \,m^{2}r +\mathrm{i}m \,q^{2}a \right)\cos \! \left(\theta \right)-\frac{m^{2}r^{2}}{2}+m \,q^{2}r -\frac{q^{4}}{2}\right)q^{4}}{\left(r^{2}+a^{2}\cos \! \left(\theta \right)^{2}\right)^{6}\left(r-ia \cos(\theta)\right)^4},\\
&\mathbb{M}_2=-\frac{8\left(\cos \! \left(\theta \right)^{2}a^{2}m^{2}+m^{2}r^{2}-2m r \,q^{2}+q^{4}\right)q^{4}\left(\mathrm{i}\cos \! \left(\theta \right)a m +m r -q^{2}\right)}{\left(r^{2}+a^{2}\cos \! \left(\theta \right)^{2}\right)^{9}\left(a^{2}\cos \! \left(\theta \right)^{2}+2\,\mathrm{i}\cos \! \left(\theta \right)a r -r^{2}\right)}.
\end{align}

We note that, for vanishing cosmological constant, $\Lambda=0$, Equations (\ref{syzyg1})-(\ref{syzvier}), reduce correctly to the syzygies found in \cite{overduincwh} for the Kerr-Newman metric.

\section{Results for the ZM curvature invariants for accelerating Kerr-Newman black holes in (anti-)de Sitter spacetime}\label{Beschleunigung}

\subsection{Accelerating and rotating charged black holes with non-zero cosmological constant $\Lambda$}

 The Pleba\'{n}ski-Demia\'{n}ski metric covers a large family of solutions which include  the  physically most significant case: that of an accelerating, rotating and charged black hole with a non-zero cosmological constant \cite{PlebanskiDemianski}. We focus on the following metric that describes an accelerating Kerr-Newman black hole in (anti-)de Sitter spacetime \cite{GrifPod},\cite{PodolskyGrif}:
 \begin{align}
 \mathrm{d}s^{2}  & =\frac{1}{\Omega^2}\Biggl\{-\frac{Q}{\rho^2}\left[\mathrm{d}t-a\sin^2\theta\mathrm{d}\phi\right]^2+
 \frac{\rho^2}{Q}\mathrm{d}r^2+\frac{\rho^2}{P}\mathrm{d}\theta^2\nonumber \\
 &+\frac{P}{\rho^2}\sin^2\theta\left[a\mathrm{d}t-(r^2+a^2)\mathrm{d}\phi\right]^2\Biggr\},
 \label{epitaxmelaniopi}
 \end{align}
 where
 \begin{align}
 \Omega=&1-\alpha r \cos\theta,\\
 P=&1-2\alpha m \cos\theta+(\alpha^2(a^2+q^2)+\frac{1}{3}\Lambda a^2)\cos^2\theta,\\
 Q=&((a^2+q^2)-2mr+r^2)(1-\alpha^2r^2)\nonumber \\
 &-\frac{1}{3}\Lambda(a^2+r^2)r^2,
 \end{align}
 and $\alpha$ is the acceleration of the black hole.

 The metric (\ref{epitaxmelaniopi}) becomes singular at the roots of $\Omega, \rho^2,Q,P$. Some of them are pseudosingularities (mere coordinate singularities) while others are true (curvature) singularities detected by the curvature invariants.
 We shall discuss the influence of the acceleration parameter $\alpha$ on these singularities.

 $\Omega$ becomes zero if:
 \begin{equation}
 r=\frac{1}{\alpha\cos\theta}.
 \label{ConformalBound}
 \end{equation}
As the metric blows up if $\Omega\rightarrow 0$, Eq. (\ref{ConformalBound}) determines the boundary of the spacetime, thus we have to restrict to regions where $\Omega>0$.
For $\alpha=0$ there is no restriction because $\Omega=1$.

$\rho^2$ becomes zero at the ring singularity:
\begin{equation}
r=0\;\;\;\text{and}\;\;\;\cos\theta=0
\end{equation}
The ring singularity $r=0,\theta=\pi/2$ is a curvature singularity for ($m\not =0$) and is unaffected by $\alpha$.

The real roots of $Q$, yield coordinate singularities which correspond to the up to $4$ horizons of the spacetime. We investigate these pseudosingularities in more detail in section \ref{orizontesgegonotwn} (see figures 7-9).

In general, at the roots of $P$ would be coordinate singularities, too. These would indicate further horizons where the vector field $\partial_{\theta}$  would change its causal character, just as the vector field $\partial_r$ does at the roots of $Q$. Nevertheless, since these horizons would lie on cones $\theta={\rm constant}$ instead of on spheres $r={\rm constant}$ would be hardly of any physical relevance \cite{arneGrezen}.  The equation $P=0$ is a quadratic equation for $\cos(\theta)$:
\begin{align}
P=0\Leftrightarrow a_1\cos^2(\theta)+b_1 \cos(\theta)+c_1=0,
\label{polarexpress}
\end{align}
where $a_1\equiv(\alpha^2(a^2+q^2)+\frac{1}{3}\Lambda a^2),b_1\equiv-2\alpha m,c_1=1$.
The roots of the quadratic equation are given by the formula:
\begin{equation}
\cos\theta_{\pm}=\frac{-b_1\pm\sqrt{b_1^2-4 a_1}}{2 a_1}.
\label{polarhorizons}
\end{equation}
If the radicand (i.e. the discriminant) in (\ref{polarhorizons}) is negative $P\not =0$ is guaranteed $\forall \theta\in\mathbb{R}$. In fact in this case, for $\Lambda>0$ we have $P>0$ for $\theta\in[0,\pi]$, since the leading coefficient in (\ref{polarexpress}) is positive.
For positive radicand (discriminant) from the theory of quadratic algebraic equations we know that $P$ has the opposite sign of $a_1$ for values of $\cos\theta$ between the two real roots and has the same sign as the sign of the leading coefficient $a_1$ for values of $\cos\theta$ outside the two roots
\footnote{Nevertheless for the values of the physical black hole parameters we investigated the real roots of $P$ occur for $\theta\not\in\mathbb{R}$.}.

 For $\Lambda=0$, if $m^2\geq a^2+q^2$ the expression for $Q$ factorises as:
 \begin{equation}
 Q=(r_{-}-r)(r_{+}-r)(1-\alpha^2 r^2),
 \end{equation}
 where
 \begin{equation}
 r_{\pm}=m\pm\sqrt{m^2-a^2-q^2}.
 \end{equation}
 The expressions for the radii $r_{\pm}$ are identical to those for the location of the outer and inner horizons of the nonaccelerating Kerr-Newman black hole. However, in the present case there is another horizon at $r=\alpha^{-1}$ known in the context of the $C-$ metric as an acceleration horizon.
 When $\Lambda \not =0$, the location of all horizons is modified.

\subsection{Computation of the ZM curvature invariants for accelerating charged and rotating black holes with $\Lambda\not =0$}

We first compute the Chern-Pontryagin invariant $K_2$ for an accelerating Kerr-Newman black hole in (anti-)de Sitter spacetime. The analytic explicit result for this fundamental invariant is:
\begin{align}
K_2&=\frac{96a\left(\alpha r \cos \! \left(\theta \right)-1\right)^{6}}{\left(r^{2}+a^{2}\cos \! \left(\theta \right)^{2}\right)^{6}}\Biggl(\cos \! \left(\theta \right)^{3}a^{4}\alpha m +\cos \! \left(\theta \right)^{3}a^{2}\alpha q^{2}r \nonumber \\
&-3\cos \! \left(\theta \right)a^{2}\alpha m \,r^{2}-\cos \! \left(\theta \right)\alpha q^{2}r^{3}-3a^{2}\cos \! \left(\theta \right)^{2}m r +a^{2}\cos \! \left(\theta \right)^{2}q^{2}+m \,r^{3}-q^{2}r^{2}\Biggr) \nonumber \\
 &\times \left(3\cos \! \left(\theta \right)^{2}a^{2}\alpha m r +2\cos \! \left(\theta \right)^{2}\alpha q^{2}r^{2}+\cos \! \left(\theta \right)^{3}a^{2}m -\alpha r^{3}m -3\cos \! \left(\theta \right)m \,r^{2}+2\cos \! \left(\theta \right)q^{2}r \right)\label{CPBeschleunigungKNdS}
\end{align}

Subsequently, we compute explicit algebraic expressions for the rest of curvature invariants from the ZM set, for the metric (\ref{epitaxmelaniopi}), with Maple\textsuperscript{TM}2021\footnote{The symbolic computation despite the formidable complexity of the tensorial operations involved runs smoothly in a modern 8GB RAM laptop whereas is particularly demanding  for the mixed invariants.}. Our results are given below:

\begin{align}
I_1&=\frac{1}{\left(r^{2}+a^{2}\cos \! \left(\theta \right)^{2}\right)^{6}}48\Biggl(\left(q^{2}r \alpha +a m \left(a \alpha -1\right)\right)a^{2}\cos \! \left(\theta \right)^{3}+\left(-2a \alpha q^{2}r^{2}-3a^{2}m \left(a \alpha +1\right)r +a^{2}q^{2}\right)\cos \! \left(\theta \right)^{2}\nonumber \\
&+\left(-\alpha q^{2}r^{3}-3a m \left(a \alpha -1\right)r^{2}-2a \,q^{2}r \right)\cos \! \left(\theta \right)+\left(m \left(a \alpha +1\right)r -q^{2}\right)r^{2}\Biggr)\nonumber \\
&\times\Biggl(a^{2}\left(q^{2}r \alpha +a m \left(a \alpha +1\right)\right)\cos \! \left(\theta \right)^{3}+\left(2a \alpha q^{2}r^{2}+3a^{2}m \left(a \alpha -1\right)r +a^{2}q^{2}\right)\cos \! \left(\theta \right)^{2}\nonumber \\
&+\left(-\alpha q^{2}r^{3}-3a m \left(a \alpha +1\right)r^{2}+2a \,q^{2}r \right)\cos \! \left(\theta \right)-\left(m \left(a \alpha -1\right)r +q^{2}\right)r^{2}\Biggr)\left(\alpha r \cos \! \left(\theta \right)-1\right)^{6}\label{AcceBeschlKNdSein},\\
I_3&=-\frac{96\left(\alpha r \cos \! \left(\theta \right)-1\right)^{9}}{\left(r^{2}+a^{2}\cos \! \left(\theta \right)^{2}\right)^{9}}\Biggl(a^{4}\left(q^{4}r^{2}\alpha^{2}+2a^{2}m \,q^{2}r \alpha^{2}+a^{2}m^{2}\left(a^{2}\alpha^{2}-3\right)\right)\cos \! \left(\theta \right)^{6}\nonumber \\
&-24\left(\frac{3m \,q^{2}r^{2}}{4}+\left(a^{2}m^{2}-\frac{q^{4}}{12}\right)r -\frac{a^{2}m \,q^{2}}{12}\right)a^{4}\alpha \cos \! \left(\theta \right)^{5}+\Biggl[-14a^{2}\alpha^{2}q^{4}r^{4}-44a^{4}m \,q^{2}\alpha^{2}r^{3}\nonumber \\
&+\left(-33a^{6}\alpha^{2}m^{2}+27a^{4}m^{2}\right)r^{2}-18a^{4}m \,q^{2}r +a^{4}q^{4}\Biggr]\cos \! \left(\theta \right)^{4}\nonumber \\
&+80r^{2}\left(\frac{11m \,q^{2}r^{2}}{20}+\left(a^{2}m^{2}-\frac{7q^{4}}{20}\right)r -\frac{11a^{2}m \,q^{2}}{20}\right)a^{2}\alpha \cos \! \left(\theta \right)^{3}\nonumber \\
&+\left(q^{4}\alpha^{2}r^{6}+18a^{2}\alpha^{2}m \,q^{2}r^{5}+\left(27a^{4}\alpha^{2}m^{2}-33a^{2}m^{2}\right)r^{4}+44a^{2}m \,q^{2}r^{3}-14a^{2}q^{4}r^{2}\right)\cos \! \left(\theta \right)^{2}\nonumber \\
&-24r^{4}\left(\frac{m \,q^{2}r^{2}}{12}+\left(a^{2}m^{2}-\frac{q^{4}}{12}\right)r -\frac{3a^{2}m \,q^{2}}{4}\right)\alpha \cos \! \left(\theta \right)\nonumber \\
&+\left(-3a^{2}\alpha^{2}m^{2}+m^{2}\right)r^{6}-2m \,q^{2}r^{5}+q^{4}r^{4}\Biggr)\nonumber \\
&\times \left(a^{2}\alpha \left(a^{2}m +q^{2}r \right)\cos \! \left(\theta \right)^{3}+\left(-3a^{2}m r +a^{2}q^{2}\right)\cos \! \left(\theta \right)^{2}+\left(-3a^{2}\alpha m \,r^{2}-\alpha q^{2}r^{3}\right)\cos \! \left(\theta \right)+m \,r^{3}-q^{2}r^{2}\right)
\end{align}

\begin{align}
I_4&=\frac{864\left(\alpha r \cos \! \left(\theta \right)-1\right)^{9}a}{\left(r^{2}+a^{2}\cos \! \left(\theta \right)^{2}\right)^{9}} \left(\frac{\cos \left(\theta \right)^{3}a^{2}m}{3}+r \alpha \left(a^{2}m +\frac{2q^{2}r}{3}\right)\cos \! \left(\theta \right)^{2}+\left(-m \,r^{2}+\frac{2}{3}q^{2}r \right)\cos \! \left(\theta \right)-\frac{\alpha r^{3}m}{3}\right)\,\nonumber \\
&\times \Biggl[a^{4}\left(\alpha^{2}q^{4}r^{2}+2a^{2}\alpha^{2}m \,q^{2}r +a^{2}m^{2}\left(a^{2}\alpha^{2}-\frac{1}{3}\right)\right)\cos \! \left(\theta \right)^{6}-8a^{4}\Biggl(\frac{11m \,q^{2}r^{2}}{12}+\left(a^{2}m^{2}-\frac{q^{4}}{4}\right)r \nonumber \\
&-\frac{a^{2}m \,q^{2}}{4}\Biggr)\alpha \cos \! \left(\theta \right)^{5}+\Biggl(-\frac{10a^{2}\alpha^{2}q^{4}r^{4}}{3}-12a^{4}\alpha^{2}m \,q^{2}r^{3}+\left(-9a^{6}\alpha^{2}m^{2}+11a^{4}m^{2}\right)r^{2}\nonumber \\
&-\frac{22a^{4}m \,q^{2}r}{3}+a^{4}q^{4}\Biggr)\cos \! \left(\theta \right)^{4}+\frac{80r^{2}a^{2}\left(\frac{9m \,q^{2}r^{2}}{20}+\left(a^{2}m^{2}-\frac{q^{4}}{4}\right)r -\frac{9a^{2}m \,q^{2}}{20}\right)\alpha \cos \left(\theta \right)^{3}}{3}+\Biggl(\alpha^{2}q^{4}r^{6}\nonumber \\
&+\frac{22a^{2}\alpha^{2}m \,q^{2}r^{5}}{3}+\left(11a^{4}\alpha^{2}m^{2}-9a^{2}m^{2}\right)r^{4}+12a^{2}m \,q^{2}r^{3}-\frac{10a^{2}q^{4}r^{2}}{3}\Biggr)\cos \! \left(\theta \right)^{2}-8r^{4}\alpha \Biggl(\frac{m \,q^{2}r^{2}}{4}\nonumber \\
&+\left(a^{2}m^{2}-\frac{q^{4}}{4}\right)r -\frac{11a^{2}m \,q^{2}}{12}\Biggr)\cos \! \left(\theta \right)-\frac{r^{4}\left(m^{2}\left(a^{2}\alpha^{2}-3\right)r^{2}+6m \,q^{2}r -3q^{4}\right)}{3}\Biggr],\\
I_5&=4\Lambda\label{constcurvaaccel},\\
I_6&=\frac{4}{\left(r^{2}+a^{2}\cos \! \left(\theta \right)^{2}\right)^{4}}\Biggl(\cos \! \left(\theta \right)^{8}\alpha^{8}q^{4}r^{8}-8\cos \! \left(\theta \right)^{7}\alpha^{7}q^{4}r^{7}+28\cos \! \left(\theta \right)^{6}\alpha^{6}q^{4}r^{6}\nonumber \\
&-56\cos \! \left(\theta \right)^{5}\alpha^{5}q^{4}r^{5}+\Lambda^{2}\cos \! \left(\theta \right)^{8}a^{8}+4\Lambda^{2}\cos \! \left(\theta \right)^{6}a^{6}r^{2}+70\cos \! \left(\theta \right)^{4}\alpha^{4}q^{4}r^{4}+6\Lambda^{2}\cos \! \left(\theta \right)^{4}a^{4}r^{4}\nonumber \\
&-56\cos \! \left(\theta \right)^{3}\alpha^{3}q^{4}r^{3}+4\Lambda^{2}\cos \! \left(\theta \right)^{2}a^{2}r^{6}+\Lambda^{2}r^{8}+28\cos \! \left(\theta \right)^{2}\alpha^{2}q^{4}r^{2}-8\cos \! \left(\theta \right)\alpha q^{4}r +q^{4}\Biggr)\label{riccisquare},\\
I_7&=\frac{4}{\left(r^{2}+a^{2}\cos \! \left(\theta \right)^{2}\right)^{4}}\Biggl(3\cos \! \left(\theta \right)^{8}\alpha^{8}q^{4}r^{8}-24\cos \! \left(\theta \right)^{7}\alpha^{7}q^{4}r^{7}+84\cos \! \left(\theta \right)^{6}\alpha^{6}q^{4}r^{6}-168\cos \! \left(\theta \right)^{5}\alpha^{5}q^{4}r^{5}\label{triaricci}\nonumber \\
& +\Lambda^{2}\cos \! \left(\theta \right)^{8}a^{8}+4\Lambda^{2}\cos \! \left(\theta \right)^{6}a^{6}r^{2}+210\cos \! \left(\theta \right)^{4}\alpha^{4}q^{4}r^{4}+6\Lambda^{2}\cos \! \left(\theta \right)^{4}a^{4}r^{4}\nonumber \\
&-168\cos \! \left(\theta \right)^{3}\alpha^{3}q^{4}r^{3}+4\Lambda^{2}\cos \! \left(\theta \right)^{2}a^{2}r^{6}+\Lambda^{2}r^{8}+84\cos \! \left(\theta \right)^{2}\alpha^{2}q^{4}r^{2}-24\cos \! \left(\theta \right)\alpha q^{4}r +3q^{4}\Biggr)\Lambda,\\
I_9&=\frac{16q^{4}}{\left(r^{2}+a^{2}\cos \! \left(\theta \right)^{2}\right)^{7}}\left(\alpha r \cos \! \left(\theta \right)-1\right)^{11}\Biggl(\cos \! \left(\theta \right)^{3}a^{4}\alpha m +\cos \! \left(\theta \right)^{3}a^{2}\alpha q^{2}r \nonumber \\
&-3\cos \! \left(\theta \right)a^{2}\alpha m \,r^{2}-\cos \! \left(\theta \right)\alpha q^{2}r^{3}-3a^{2}\cos \! \left(\theta \right)^{2}m r +a^{2}\cos \! \left(\theta \right)^{2}q^{2}+m \,r^{3}-q^{2}r^{2}\Biggr)\label{mixerI9accel},
\end{align}

\begin{align}
&I_8=\frac{1}{\left(r^{2}+a^{2}\cos \! \left(\theta \right)^{2}\right)^{8}}\nonumber \\
&\times\Biggl(\left(4\alpha^{16}q^{8}r^{16}+24\Lambda^{2}a^{8}\alpha^{8}q^{4}r^{8}+4\Lambda^{4}a^{16}\right)\cos \! \left(\theta \right)^{16}+\left(-64\alpha^{15}q^{8}r^{15}-192\Lambda^{2}a^{8}\alpha^{7}q^{4}r^{7}\right)\cos \! \left(\theta \right)^{15}\nonumber \\
&+32r^{2}\left(15\alpha^{14}q^{8}r^{12}+3\Lambda^{2}a^{6}\alpha^{8}q^{4}r^{8}+21\Lambda^{2}a^{8}\alpha^{6}q^{4}r^{4}+\Lambda^{4}a^{14}\right)\cos \! \left(\theta \right)^{14}-1344r^{5}\alpha^{5}q^{4}\Biggl[\frac{5}{3}q^{4}r^{8}\alpha^{8}+\frac{4}{7}a^{6}r^{4}\alpha^{2}\Lambda^{2}\nonumber \\
&+a^{8}\Lambda^{2}\Biggr]\cos \! \left(\theta \right)^{13}+112r^{4}\left(\left(65q^{8}\alpha^{12}+\frac{9}{7}a^{4}q^{4}\alpha^{8}\Lambda^{2}\right)r^{8}+24a^{6}q^{4}r^{4}\alpha^{6}\Lambda^{2}+a^{8}\Lambda^{2}\left(15q^{4}\alpha^{4}+a^{4}\Lambda^{2}\right)\right)\cos \! \left(\theta \right)^{12}\nonumber \\
&-1344r^{3}\alpha^{3}\left(\left(13q^{4}\alpha^{8}+\frac{6}{7}a^{4}\alpha^{4}\Lambda^{2}\right)r^{8}+4a^{6}r^{4}\alpha^{2}\Lambda^{2}+a^{8}\Lambda^{2}\right)q^{4}\cos \! \left(\theta \right)^{11}+224r^{2}\Biggl[\frac{3\Lambda^{2}a^{2}\alpha^{8}q^{4}r^{12}}{7}\nonumber \\
&+\left(143q^{8}\alpha^{10}+18\Lambda^{2}a^{4}\alpha^{6}q^{4}\right)r^{8}+a^{6}\Lambda^{2}\left(30q^{4}\alpha^{4}+a^{4}\Lambda^{2}\right)r^{4}+3a^{8}q^{4}\alpha^{2}\Lambda^{2}\Biggr]\cos \! \left(\theta \right)^{10}-192r \alpha q^{4}\Biggl[4a^{2}r^{12}\alpha^{6}\Lambda^{2}\nonumber \\
&+\left(\frac{715}{3}q^{4}\alpha^{8}+42a^{4}\alpha^{4}\Lambda^{2}\right)r^{8}+28a^{6}r^{4}\alpha^{2}\Lambda^{2}+a^{8}\Lambda^{2}\Biggr]\cos \! \left(\theta \right)^{9}+\Biggl[24\Lambda^{2}\alpha^{8}q^{4}r^{16}+2688\Lambda^{2}a^{2}\alpha^{6}q^{4}r^{12}\nonumber \\
&+\left(51480q^{8}\alpha^{8}+10080\Lambda^{2}a^{4}\alpha^{4}q^{4}+280\Lambda^{4}a^{8}\right)r^{8}+2688\Lambda^{2}a^{6}\alpha^{2}q^{4}r^{4}+24\Lambda^{2}a^{8}q^{4}\Biggr]\cos \! \left(\theta \right)^{8}\nonumber \\
&-768r^{3}\alpha q^{4}\left(\frac{r^{12}\alpha^{6}\Lambda^{2}}{4}+7a^{2}r^{8}\alpha^{4}\Lambda^{2}+\left(\frac{715}{12}q^{4}\alpha^{6}+\frac{21}{2}a^{4}\alpha^{2}\Lambda^{2}\right)r^{4}+a^{6}\Lambda^{2}\right)\cos \! \left(\theta \right)^{7}+96r^{2}\Biggl[7q^{4}r^{12}\alpha^{6}\Lambda^{2}\nonumber \\
&+\left(70a^{2}q^{4}\alpha^{4}\Lambda^{2}+\frac{7}{3}a^{6}\Lambda^{4}\right)r^{8}+\left(\frac{1001}{3}q^{8}\alpha^{6}+42a^{4}q^{4}\alpha^{2}\Lambda^{2}\right)r^{4}+a^{6}q^{4}\Lambda^{2}\Biggr]\cos \! \left(\theta \right)^{6}-1152r^{5}\alpha \Biggl[\frac{7}{6}r^{8}\alpha^{4}\Lambda^{2}\nonumber \\
&+\frac{14}{3}a^{2}r^{4}\alpha^{2}\Lambda^{2}+\frac{91}{6}q^{4}\alpha^{4}+a^{4}\Lambda^{2}\Biggr]q^{4}\cos \! \left(\theta \right)^{5}+\Biggl(1680\Lambda^{2}\alpha^{4}q^{4}r^{12}+112\Lambda^{4}a^{4}r^{12}+2688\Lambda^{2}a^{2}\alpha^{2}q^{4}r^{8}
\nonumber \\
&+7280\alpha^{4}q^{8}r^{4}+144\Lambda^{2}a^{4}q^{4}r^{4}\Biggr)\cos \! \left(\theta \right)^{4}-768r^{3}\left(\frac{7}{4}r^{8}\alpha^{2}\Lambda^{2}+a^{2}r^{4}\Lambda^{2}+\frac{35}{12}q^{4}\alpha^{2}\right)\alpha q^{4}\cos \! \left(\theta \right)^{3}\nonumber \\
&+96r^{2}\left(\frac{1}{3}\Lambda^{4}a^{2}r^{12}+7q^{4}r^{8}\alpha^{2}\Lambda^{2}+\Lambda^{2}a^{2}q^{4}r^{4}+5q^{8}\alpha^{2}\right)\cos \! \left(\theta \right)^{2}-64q^{4}r \alpha \left(3\Lambda^{2}r^{8}+q^{4}\right)\cos \! \left(\theta \right)+4\Lambda^{4}r^{16}\nonumber \\
&+24\Lambda^{2}q^{4}r^{8}+4q^{8}\Biggr)\label{achtricci},\\
&I_{10}=-\frac{16 a q^{4} \left(\alpha r \cos \! \left(\theta \right)-1\right)^{11}}{\left(r^{2}+a^{2}\cos \! \left(\theta \right)^{2}\right)^{7}}\Biggl(3\cos \! \left(\theta \right)^{2}a^{2}\alpha m r +2\cos \! \left(\theta \right)^{2}\alpha q^{2}r^{2}+\cos \! \left(\theta \right)^{3}a^{2}m \nonumber \\
&-\alpha r^{3}m -3\cos \! \left(\theta \right)m \,r^{2}+2\cos \! \left(\theta \right)q^{2}r \Biggr),
\end{align}

\begin{align}
I_{11}&=\frac{64q^{4}\left(\alpha r \cos \! \left(\theta \right)-1\right)^{14}}{\left(r^{2}+a^{2}\cos \! \left(\theta \right)^{2}\right)^{10}}\Biggl[\left(q^{2}r \alpha +a m \left(a \alpha +1\right)\right)a^{2}\cos \! \left(\theta \right)^{3}+\left(2a \,q^{2}\alpha r^{2}+3a^{2}m \left(a \alpha -1\right)r +a^{2}q^{2}\right)\cos \! \left(\theta \right)^{2}\nonumber \\
&+\left(-\alpha q^{2}r^{3}-3a m \left(a \alpha +1\right)r^{2}+2a \,q^{2}r \right)\cos \! \left(\theta \right)-r^{2}\left(m \left(a \alpha -1\right)r +q^{2}\right)\Biggr]\nonumber \\
&\times \Biggl(a^{2}\left(q^{2}r \alpha +a m \left(a \alpha -1\right)\right)\cos \! \left(\theta \right)^{3}+\left(-2a \,q^{2}\alpha r^{2}-3a^{2}m \left(a \alpha +1\right)r +a^{2}q^{2}\right)\cos \! \left(\theta \right)^{2}\nonumber \\
&+\left(-\alpha q^{2}r^{3}-3a m \left(a \alpha -1\right)r^{2}-2a \,q^{2}r \right)\cos \! \left(\theta \right)+\left(m \left(a \alpha +1\right)r -q^{2}\right)r^{2}\Biggr),\\
I_{12}&=-\frac{128 a q^{4}\left(\alpha r \cos \! \left(\theta \right)-1\right)^{14}}{\left(r^{2}+a^{2}\cos \! \left(\theta \right)^{2}\right)^{10}} \Biggl(3\cos \! \left(\theta \right)^{2}a^{2}\alpha m r +2\cos \! \left(\theta \right)^{2}\alpha q^{2}r^{2}+\cos \! \left(\theta \right)^{3}a^{2}m \nonumber\\
&  -\alpha r^{3}m -3\cos \! \left(\theta \right)m \,r^{2}+2\cos \! \left(\theta \right)q^{2}r \Biggr)\nonumber \\
&\times \Biggl[\cos \! \left(\theta \right)^{3}a^{4}\alpha m +\cos \! \left(\theta \right)^{3}a^{2}\alpha q^{2}r -3\cos \! \left(\theta \right)a^{2}\alpha m \,r^{2}-\cos \! \left(\theta \right)\alpha q^{2}r^{3}-3a^{2}\cos \! \left(\theta \right)^{2}m r \nonumber \\
&+a^{2}\cos \! \left(\theta \right)^{2}q^{2}+m \,r^{3}-q^{2}r^{2}\Biggr],\\
I_{15}&=\frac{4q^{4}}{\left(r^{2}+a^{2}\cos \! \left(\theta \right)^{2}\right)^{8}}\Biggl(\left(\alpha^{2}q^{4}r^{2}+2a^{2}\alpha^{2}m \,q^{2}r +a^{2}m^{2}\left(a^{2}\alpha^{2}+1\right)\right)\cos \! \left(\theta \right)^{2}\nonumber \\
&+ 2q^{2}\alpha \left(a^{2}m -m \,r^{2}+q^{2}r \right)\cos \! \left(\theta \right)+m^{2}\left(a^{2}\alpha^{2}+1\right)r^{2}-2m \,q^{2}r +q^{4}\Biggr)
\left(\alpha r \cos \! \left(\theta \right)-1\right)^{14}, \\
I_{16}&=-\frac{8 \left(\alpha  r \cos \! \left(\theta \right)-1\right)^{17} q^{4}
}{\left(r^{2}+a^{2} \cos \! \left(\theta \right)^{2}\right)^{11}} \Biggl[\left(\alpha^{2} q^{4} r^{2}+2 a^{2} \alpha^{2} m \,q^{2} r +a^{2} m^{2} \left(a^{2} \alpha^{2}+1\right)\right) \cos \! \left(\theta \right)^{2}\nonumber \\
&+2 q^{2} \alpha  \left(a^{2} m -m \,r^{2}+q^{2} r \right) \cos \! \left(\theta \right)+m^{2} \left(a^{2} \alpha^{2}+1\right) r^{2}-2 m r \,q^{2}+q^{4}\Biggr]\nonumber \\
&\times \Biggl(a^{2} \alpha  \left(a^{2} m +q^{2} r \right) \cos \! \left(\theta \right)^{3}+\left(-3 a^{2} m r +a^{2} q^{2}\right) \cos \! \left(\theta \right)^{2}\nonumber \\
&+\left(-3 a^{2} \alpha  m \,r^{2}-\alpha  q^{2} r^{3}\right) \cos \! \left(\theta \right)+m \,r^{3}-q^{2} r^{2}\Biggr),
\end{align}
\begin{align}
I_{17}&=\frac{24aq^{4}\left(\alpha r \cos \left(\theta \right)-1\right)^{17}}{\left(r^{2}+a^{2}\cos \! \left(\theta \right)^{2}\right)^{11}}\,\Biggl(\frac{\cos \left(\theta \right)^{3}a^{2}m}{3}+r \alpha \left(a^{2}m +\frac{2q^{2}r}{3}\right)\cos \! \left(\theta \right)^{2}\nonumber\\
&+\left(-m \,r^{2}+\frac{2}{3}q^{2}r \right)\cos \! \left(\theta \right)-\frac{\alpha r^{3}m}{3}\Biggr) \,\Biggl(\left(\alpha^{2}q^{4}r^{2}+2a^{2}\alpha^{2}m \,q^{2}r +a^{2}m^{2}\left(a^{2}\alpha^{2}+1\right)\right)\cos \! \left(\theta \right)^{2}\nonumber \\
&+2q^{2}\alpha \left(a^{2}m -m \,r^{2}+q^{2}r \right)\cos \! \left(\theta \right)+m^{2}\left(a^{2}\alpha^{2}+1\right)r^{2}-2m \,q^{2}r +q^{4}\Biggr)\label{BeschleunKNdSsiebundzehn}.
\end{align}

We summarise our results as follows:
\begin{theorem}\label{courbureinvtheoriazwei}
The exact algebraic expressions for the curvature invariants calculated for the accelerating  Kerr-Newman black hole in (anti-)de Sitter spacetime are given in Equations (\ref{AcceBeschlKNdSein})-(\ref{BeschleunKNdSsiebundzehn}) and (\ref{CPBeschleunigungKNdS}).
\end{theorem}

\begin{remark}
For vanishing acceleration of the black hole, i.e. $\alpha=0$, we recover the results of Theorem \ref{courbureinvriemein}.
\end{remark}

For completeness we present the result for the Kretschmann scalar invariant:

\begin{theorem}\label{KretschmannAccelaKNdS}
The  Kretschmann scalar invariant $K=R_{\alpha\beta\gamma\delta}R^{\alpha\beta\gamma\delta}$ for the accelerating Kerr-Newman black hole in (anti-)de Sitter spacetime is given by:
\begin{align}
K&=\frac{1}{\left(r^{2}+a^{2}\cos \! \left(\theta \right)^{2}\right)^{6}}48\Biggl(\left(q^{2}r \alpha +a m \left(a \alpha -1\right)\right)a^{2}\cos \! \left(\theta \right)^{3}+\left(-2a \alpha q^{2}r^{2}-3a^{2}m \left(a \alpha +1\right)r +a^{2}q^{2}\right)\cos \! \left(\theta \right)^{2}\nonumber \\
&+\left(-\alpha q^{2}r^{3}-3a m \left(a \alpha -1\right)r^{2}-2a \,q^{2}r \right)\cos \! \left(\theta \right)+\left(m \left(a \alpha +1\right)r -q^{2}\right)r^{2}\Biggr)\nonumber \\
&\times\Biggl(a^{2}\left(q^{2}r \alpha +a m \left(a \alpha +1\right)\right)\cos \! \left(\theta \right)^{3}+\left(2a \alpha q^{2}r^{2}+3a^{2}m \left(a \alpha -1\right)r +a^{2}q^{2}\right)\cos \! \left(\theta \right)^{2}\nonumber \\
&+\left(-\alpha q^{2}r^{3}-3a m \left(a \alpha +1\right)r^{2}+2a \,q^{2}r \right)\cos \! \left(\theta \right)-\left(m \left(a \alpha -1\right)r +q^{2}\right)r^{2}\Biggr)\left(\alpha r \cos \! \left(\theta \right)-1\right)^{6}\nonumber \\
&+\frac{8}{\left(r^{2}+a^{2}\cos \! \left(\theta \right)^{2}\right)^{4}}\Biggl(\cos \! \left(\theta \right)^{8}\alpha^{8}q^{4}r^{8}-8\cos \! \left(\theta \right)^{7}\alpha^{7}q^{4}r^{7}+28\cos \! \left(\theta \right)^{6}\alpha^{6}q^{4}r^{6}\nonumber \\
&-56\cos \! \left(\theta \right)^{5}\alpha^{5}q^{4}r^{5}+\Lambda^{2}\cos \! \left(\theta \right)^{8}a^{8}+4\Lambda^{2}\cos \! \left(\theta \right)^{6}a^{6}r^{2}+70\cos \! \left(\theta \right)^{4}\alpha^{4}q^{4}r^{4}+6\Lambda^{2}\cos \! \left(\theta \right)^{4}a^{4}r^{4}\nonumber \\
&-56\cos \! \left(\theta \right)^{3}\alpha^{3}q^{4}r^{3}+4\Lambda^{2}\cos \! \left(\theta \right)^{2}a^{2}r^{6}+\Lambda^{2}r^{8}+28\cos \! \left(\theta \right)^{2}\alpha^{2}q^{4}r^{2}-8\cos \! \left(\theta \right)\alpha q^{4}r +q^{4}\Biggr)\nonumber \\
&-\frac{16\Lambda^2}{3}.
\end{align}
\end{theorem}

We note that we have also checked our results for the curvature invariants for accelerating Kerr-Newman-(anti-)de Sitter black holes within the NP formalism , as we did in Section \ref{RogerEwa} for the case of non-accelerating Kerr-Newman-(anti-)de Sitter black holes. For instance, the only non-zero curvature scalars in the NP-formalism for the metric (\ref{epitaxmelaniopi}) are the Ricci scalars:
\begin{equation}
\Phi_{11}=\frac{1}{2}q^2\frac{(1-\alpha r\cos(\theta))^4}{(r^2+a^2 \cos(\theta)^2)^2},\;\;\;\text{and}\;\;\;\Lambda,
\end{equation}
and the Weyl scalar:
\begin{equation}
\Psi_2=-\frac{\left(1-\alpha  r \cos \! \left(\theta \right)\right)^{3} \left(m \left(i a \alpha +1\right) \left(r +i a \cos \! \left(\theta \right)\right)-q^{2} \left(1+\alpha  r \cos \! \left(\theta \right)\right)\right)}{\left(r -i a \cos \! \left(\theta \right)\right)^{3} \left(r +i a \cos \! \left(\theta \right)\right)}
\end{equation}
As a result, following the same procedure as we did in section \ref{RogerEwa},  we obtain for the curvature invariant $I_6$ the explicit compact form:
\begin{equation}
I_6=\frac{4q^{4}\left(1-\alpha r \cos \! \left(\theta \right)\right)^{8}}{\left(r^{2}+a^{2}\cos \! \left(\theta \right)^{2}\right)^{4}}+4\Lambda^{2},
\end{equation}
which is in total agreement with eqn.(\ref{riccisquare}) obtained with tensorial computation using a Maple code.
Likewise within the NP formalism we derive the following explicit algebraic expression for the curvature invariants $I_7,I_8$:
\begin{align}
&I_7=\frac{12\Lambda q^{4}\left(1-\alpha r \cos \! \left(\theta \right)\right)^{8}}{\left(r^{2}+a^{2}\cos \! \left(\theta \right)^{2}\right)^{4}}+4\Lambda^{3},\\
&I_8=\frac{4q^{8}\left(1-\alpha r \cos \! \left(\theta \right)\right)^{16}}{\left(r^{2}+a^{2}\cos \! \left(\theta \right)^{2}\right)^{8}}+4\left(\frac{12\Lambda q^{4}\left(1-\alpha r \cos \! \left(\theta \right)\right)^{8}}{\left(r^{2}+a^{2}\cos \! \left(\theta \right)^{2}\right)^{4}}+4\Lambda^{3}\right)\Lambda \nonumber \\
&-6\left(\frac{4q^{4}\left(1-\alpha r \cos \! \left(\theta \right)\right)^{8}}{\left(r^{2}+a^{2}\cos \! \left(\theta \right)^{2}\right)^{4}}+4\Lambda^{2}\right)\Lambda^{2}+12\Lambda^{4}.
\end{align}
a result that fully agrees with eqns.(\ref{triaricci}) and (\ref{achtricci}) respectively.

The syzygies Eqns(\ref{syzyg1}),(\ref{syzygzwei}),(\ref{syzdrei}),(\ref{syzvier}) that the curvature invariants satisfy are still valid in the more general case of accelerating charged and rotating black holes with a non-zero cosmological constant.

\subsubsection{Horizons of the accelerating KN(a)dS black hole}\label{orizontesgegonotwn}
It is convenient to introduce a dimensionless cosmological parameter
\begin{equation}
\Lambda=\frac{1}{3}\Lambda m^2,
\end{equation}
and set $m=1$.
Horizons of the accelerating KN(a)dS geometries are given by the condition $Q=0$, which determines pseudosingularities of the spacetime interval (\ref{epitaxmelaniopi}).
This condition yields the relations:
\begin{equation}
\Lambda=\Lambda_h(r,a,q,\alpha)=\frac{(r^2+a^2+q^2)(1-\alpha^2 r^2)}{r^2(r^2+a^2)}.
\end{equation}

\begin{figure}
[ptbh]
\begin{center}
\includegraphics[height=2.4526in, width=3.3797in ]{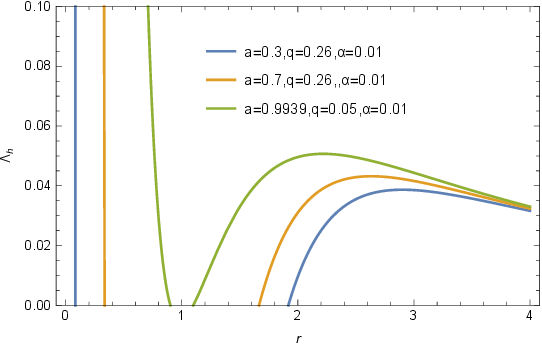}
 \caption{Horizons of the accelerating KNdS black hole as defined by the function $\Lambda_h(r;a,q,\alpha)$, for various values of the Kerr parameter and the electric charge and for $\alpha=0.01,m=1$. }%
\label{HoriEpiKNdSalp001}%
\end{center}
\end{figure}

\begin{figure}
[ptbh]
\begin{center}
\includegraphics[height=2.4526in, width=3.3797in ]{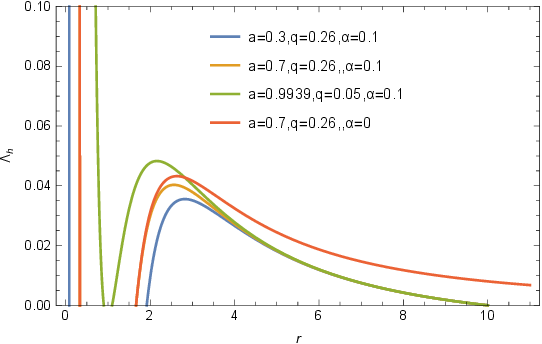}
 \caption{Horizons of the accelerating KNdS black hole as defined by the function $\Lambda_h(r;a,q,\alpha)$, for various values of the Kerr parameter $a$, the electric charge $q$ and for $\alpha=0.1,m=1$. Also the horizons for a non-accelerating KNdS black hole are displayed for the set of parameters: $a=0.7,q=0.26,\alpha=0,m=1$-red curve. }%
\label{HorizLpalp01alp0}%
\end{center}
\end{figure}

\begin{figure}
[ptbh]
\begin{center}
\includegraphics[height=2.4526in, width=3.3797in ]{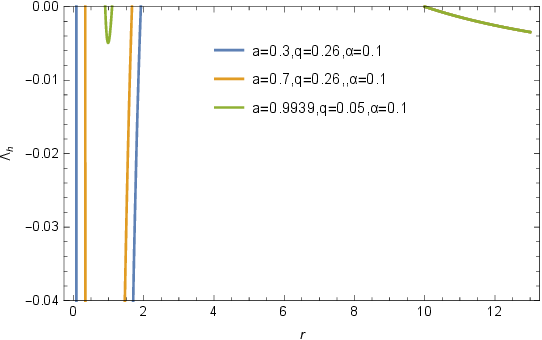}
 \caption{Horizons of the accelerating KNadS black hole as defined by the function $\Lambda_h(r;a,q,\alpha)$, for various values of the Kerr parameter $a$, the electric charge $q$ and for $\alpha=0.1,m=1$.  }%
\label{HorizLnaccel01}%
\end{center}
\end{figure}

\subsection{Exploring the geometry of spacetime surrounding the accelerating, rotating and charged black hole singularity through curvature invariants}\label{eswterikomelanisopis}

\begin{figure}
[ptbh]
\begin{center}
\includegraphics[height=2.4526in, width=3.3797in ]{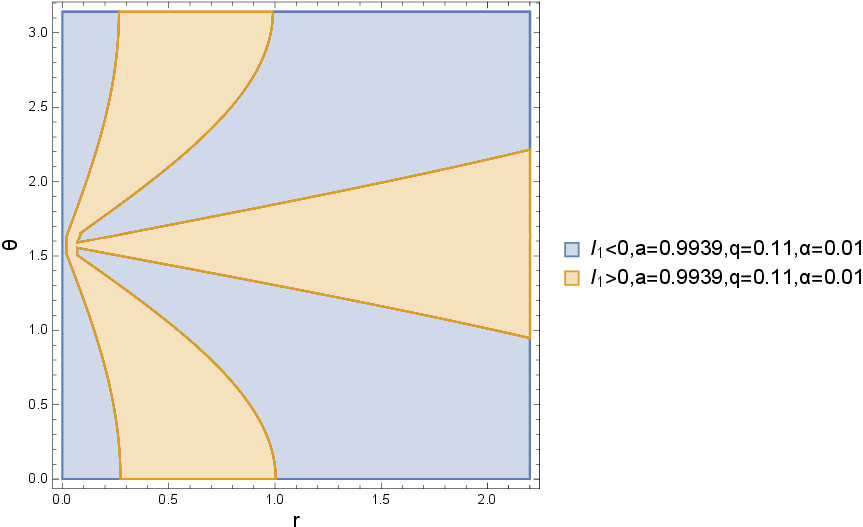}
 \caption{Regions of negative and positive $I_1$, eqn.(\ref{AcceBeschlKNdSein}) for an accelerating, charged and rotating black hole for the choice of values for the black hole parameters: $a=0.9939,q=0.11,\alpha=0.01,m=1$. }%
\label{regionI1a09939q011alp001}%
\end{center}
\end{figure}

\begin{figure}
[ptbh]
\begin{center}
\includegraphics[height=2.4526in, width=3.3797in ]{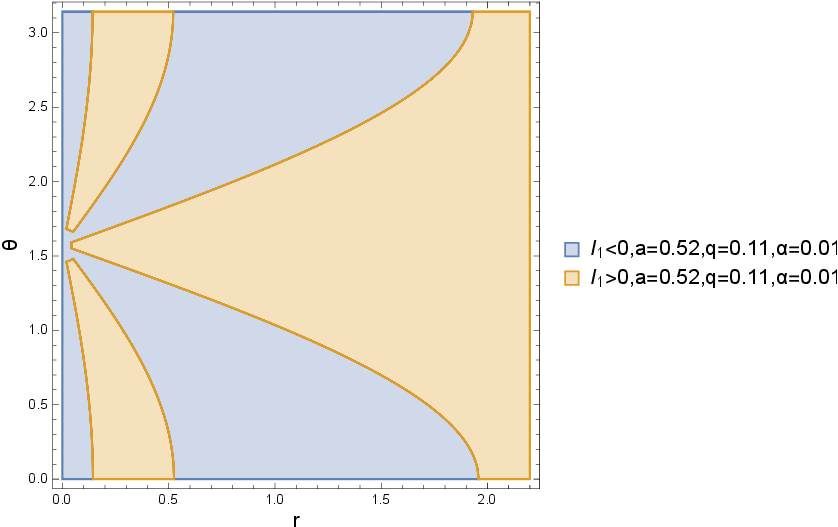}
 \caption{Regions of negative and positive $I_1$, eqn.(\ref{AcceBeschlKNdSein}) for an accelerating, charged and rotating black hole for the choice of values for the black hole parameters: $a=0.52,q=0.11,\alpha=0.01,m=1$. }%
\label{regionI1a052q011alp001}%
\end{center}
\end{figure}

\begin{figure}
[ptbh]
\begin{center}
\includegraphics[height=2.4526in, width=3.3797in ]{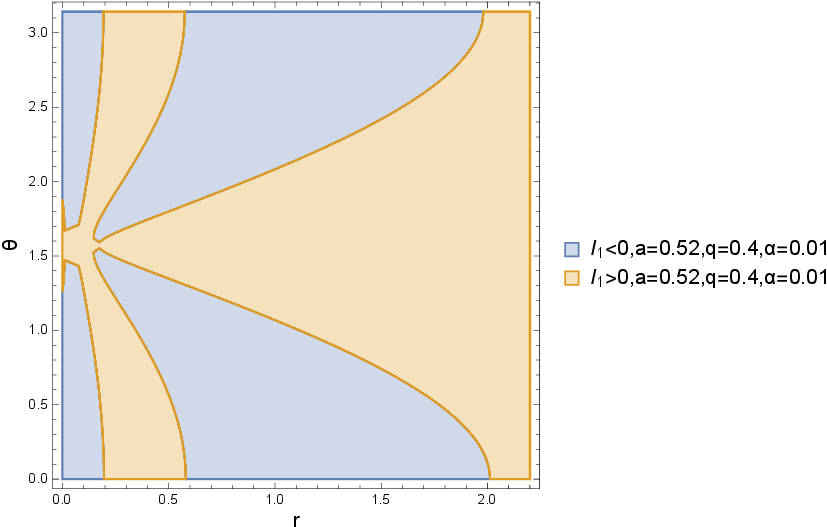}
 \caption{Regions of negative and positive $I_1$, eqn.(\ref{AcceBeschlKNdSein}) for an accelerating, charged and rotating black hole for the choice of values for the black hole parameters: $a=0.52,q=0.4,\alpha=0.01,m=1$. }%
\label{regionI1a052q04alp001}%
\end{center}
\end{figure}

\begin{figure}
[ptbh]
\begin{center}
\includegraphics[height=2.4526in, width=3.3797in ]{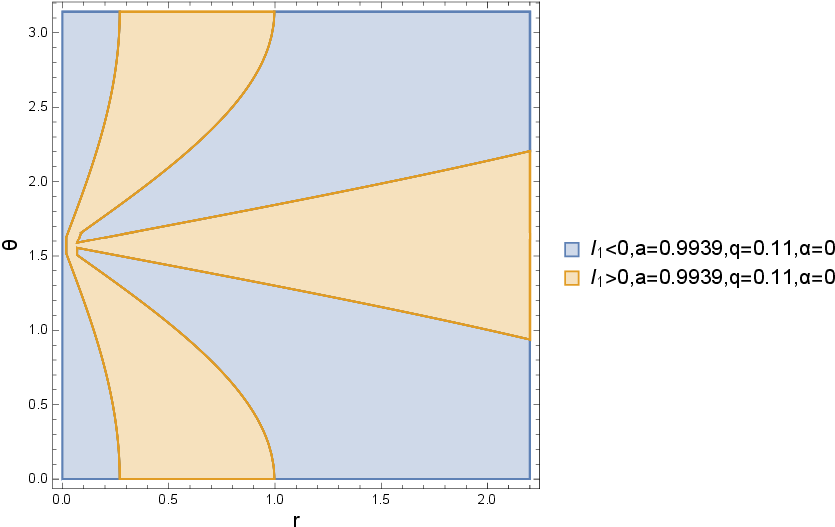}
 \caption{Regions of negative and positive $I_1$, eqn.(\ref{CIZMein}) for a  non-accelerating, charged and rotating black hole for the choice of values for the black hole parameters: $a=0.9939,q=0.11,\alpha=0,m=1$. }%
\label{regionalpha0I1a09939q011}%
\end{center}
\end{figure}

\begin{figure}
[ptbh]
\begin{center}
\includegraphics[height=2.4526in, width=3.3797in ]{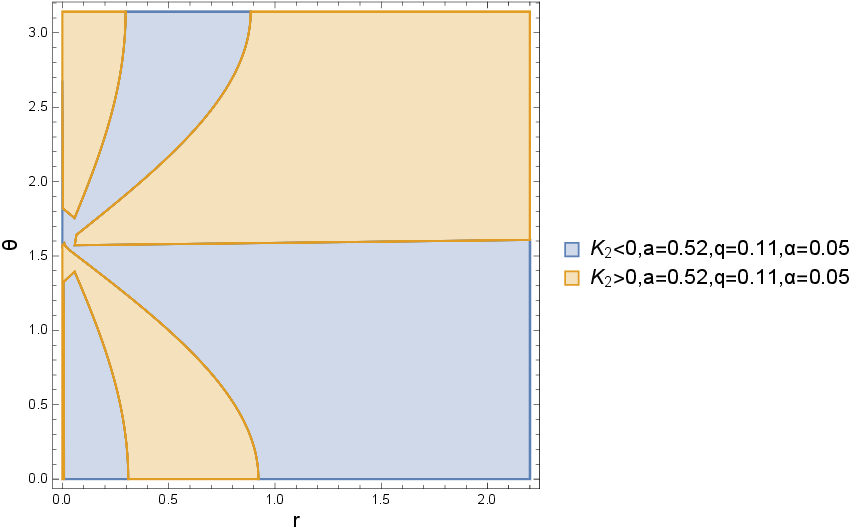}
 \caption{Regions of negative and positive Chern-Pontryagin invariant $K_2$, eqn.(\ref{CPBeschleunigungKNdS}) for an accelerating Kerr-Newman black hole in (anti-)de Sitter spacetime for the choice of values for the black hole parameters: $a=0.52,q=0.11,\alpha=0.05,m=1$. }%
\label{regionK2a052q011alp005}%
\end{center}
\end{figure}

\begin{figure}
[ptbh]
\begin{center}
\includegraphics[height=2.4526in, width=3.3797in ]{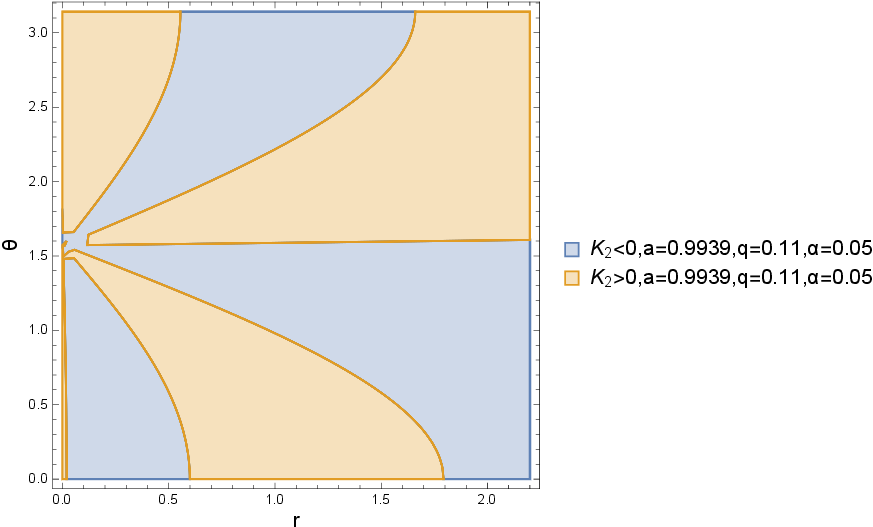}
 \caption{Regions of negative and positive Chern-Pontryagin invariant $K_2$, eqn.(\ref{CPBeschleunigungKNdS}) for an accelerating Kerr-Newman black hole in (anti-)de Sitter spacetime for the choice of values for the black hole parameters: $a=0.9939,q=0.11,\alpha=0.05,m=1$. }%
\label{regionK2a09939q011alp005}%
\end{center}
\end{figure}

\begin{figure}
[ptbh]
\begin{center}
\includegraphics[height=2.4526in, width=3.3797in ]{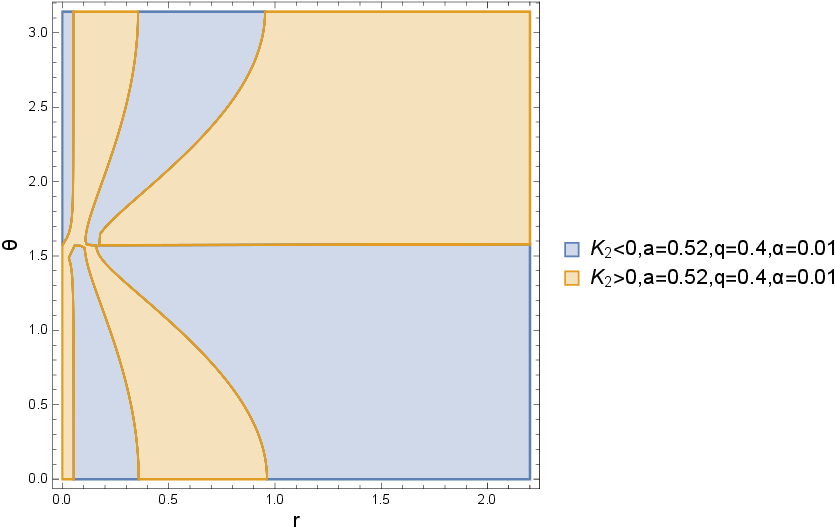}
 \caption{Regions of negative and positive Chern-Pontryagin invariant $K_2$, eqn.(\ref{CPBeschleunigungKNdS}) for an accelerating Kerr-Newman black hole in (anti-)de Sitter spacetime for the choice of values for the black hole parameters: $a=0.52,q=0.4,\alpha=0.01,m=1$. }%
\label{regionK2a052q04alp001}%
\end{center}
\end{figure}

\begin{figure}
[ptbh]
\begin{center}
\includegraphics[height=2.4526in, width=3.3797in ]{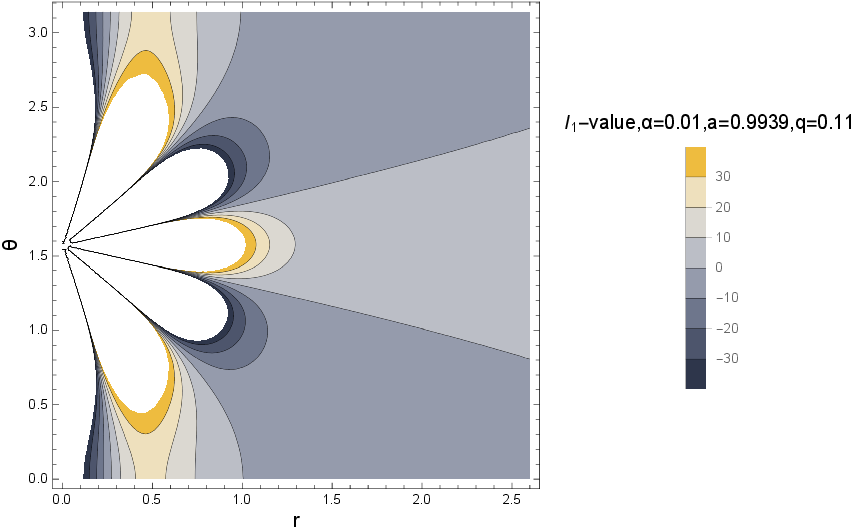}
 \caption{Contour plot of level curves of constant $I_1$ (eqn.\ref{AcceBeschlKNdSein}) for the choice of parameter values: $a=0.9939,q=0.11,\alpha=0.01,m=1$, in the $r-\theta$ plane. }%
\label{LevelcurvI1a09939q011al001}%
\end{center}
\end{figure}

\begin{figure}
[ptbh]
\begin{center}
\includegraphics[height=2.4526in, width=3.3797in ]{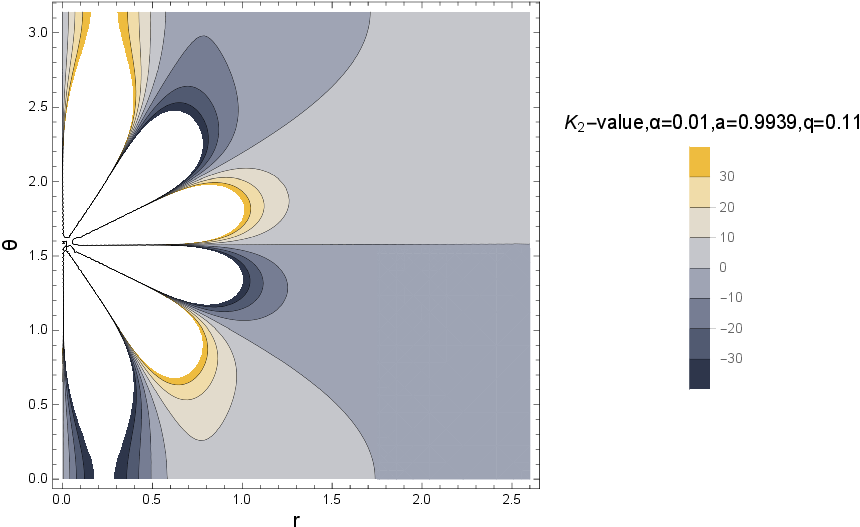}
 \caption{Contour plot of level curves of constant Chern-Pontryagin $K_2$ invariant (eqn.\ref{CPBeschleunigungKNdS}) for the choice of parameter values: $a=0.9939,q=0.11,\alpha=0.01,m=1$, in the $r-\theta$ plane. }%
\label{LevelcurvK2a09939q011al001}%
\end{center}
\end{figure}

\begin{figure}[ptbh]
\centering
  \begin{subfigure}[b]{.60\linewidth}
    \centering
    \includegraphics[width=.99\textwidth]{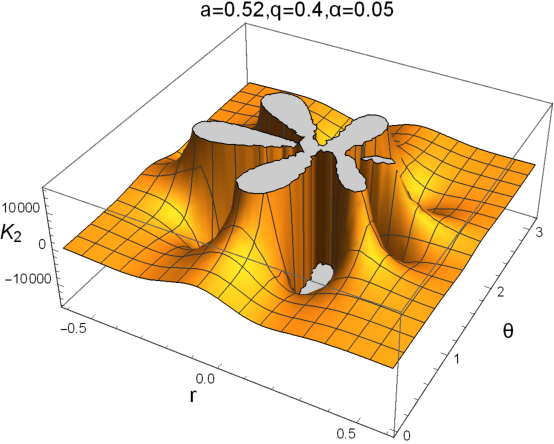}
    \caption{ 3d plot of  $K_{2}$.}\label{diag3dK2a052q04alp005}
  \end{subfigure}%
  \begin{subfigure}[b]{.60\linewidth}
    \centering
    \includegraphics[width=.99\textwidth]{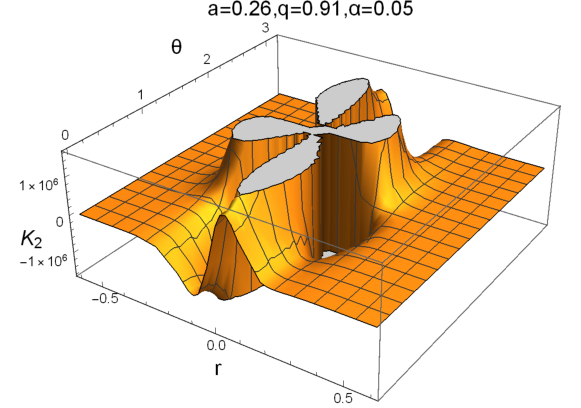}
    \caption{3d  plot of $K_{2}$}\label{diasa3K2a026q091alp005}
  \end{subfigure}\\
  \begin{subfigure}[b]{.60\linewidth}
    \centering
    \includegraphics[width=.99\textwidth]{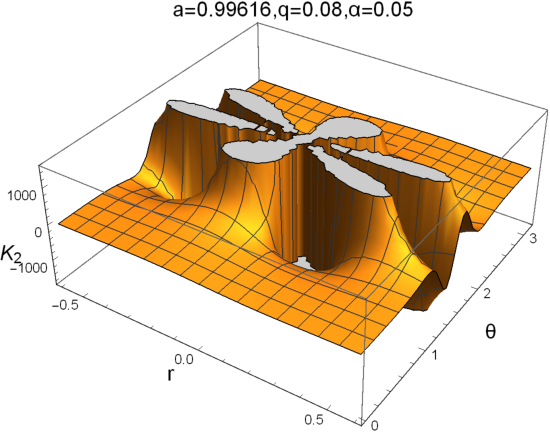}
    \caption{3d Plot of $K_{2}$.}\label{xwros3K2a099616q008alp005}
  \end{subfigure}%
  \caption{The Chern-Potryagin Weyl Invariant $K_2$,  eqn.(\ref{CPBeschleunigungKNdS}), plotted as a function of the Boyer-Lindquist coordinates $r$ and $\theta$ for an accelerating charged and rotating black hole of mass $m=1$, and the following sets of values for the physical  parameters: (a) $a=52,q=0.4,\alpha=0.05$.  (b) $a=0.26,q=0.91,\alpha=0.05$ . (c) $a=0.99616,q=0.08,\alpha=0.05$.}\label{DreiGraphChernPontryaginK2accel}
\end{figure}

\begin{figure}
[ptbh]
\begin{center}
\includegraphics[height=2.4526in, width=3.3797in ]{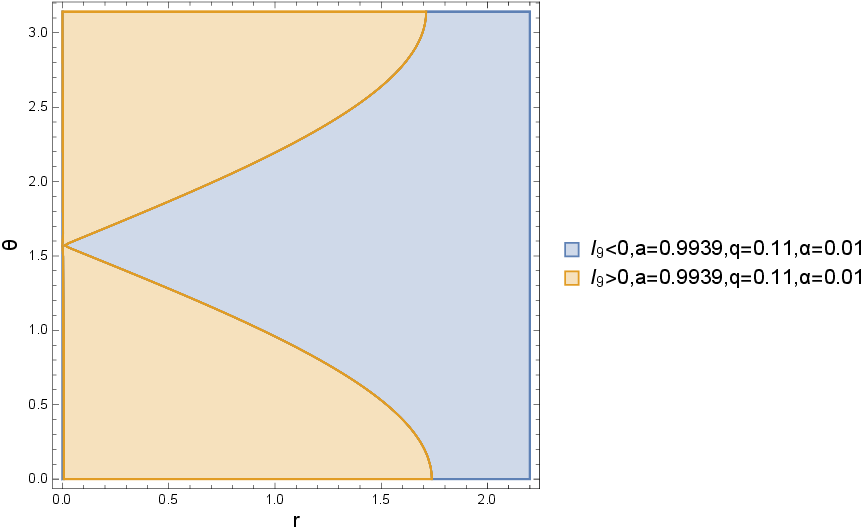}
 \caption{Regions of negative and positive mixed invariant $I_9$,  eqn.(\ref{mixerI9accel}), in the $r-\theta$ space, for an accelerating charged and rotating black hole of mass $m=1$, and physical  parameters $a=0.9939,q=0.11,\alpha=0.01$ . }%
\label{regionmI9a09939q011alp001}%
\end{center}
\end{figure}

\begin{figure}
[ptbh]
\begin{center}
\includegraphics[height=2.4526in, width=3.3797in ]{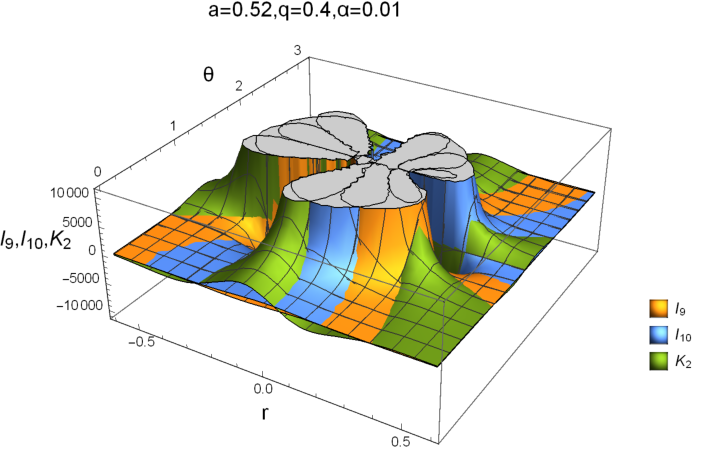}
 \caption{The curvature invariants $I_9,I_{10},K_2$,   plotted as a function of the Boyer-Lindquist coordinates $r$ and $\theta$ for an accelerating rotating and charged  black hole of mass $m=1$, and physical black hole parameters $a=0.52,q=0.4,\alpha=0.01$ . }%
\label{diag3dI9I10K2a052q04alp001}%
\end{center}
\end{figure}

\begin{figure}
[ptbh]
\begin{center}
\includegraphics[height=2.4526in, width=3.3797in ]{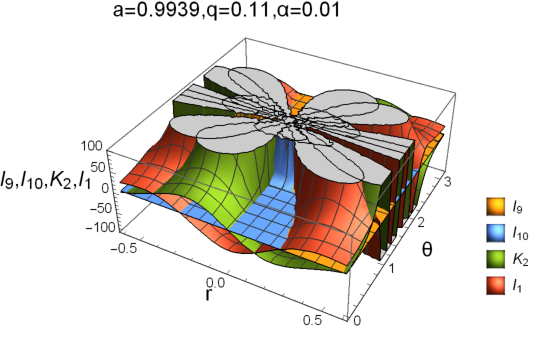}
 \caption{The curvature invariants $I_9,I_{10},K_2,I_1$,   plotted as a function of the Boyer-Lindquist coordinates $r$ and $\theta$ for an accelerating rotating and charged  black hole of mass $m=1$, and physical black hole parameters $a=0.9939,q=0.11,\alpha=0.01$ . }%
\label{diag3dI1K2I9I10a09939q011alp001}%
\end{center}
\end{figure}

In the regional plots, Figures \ref{regionI1a09939q011alp001}-\ref{regionalpha0I1a09939q011},\ref{regionK2a052q011alp005}-\ref{regionK2a052q04alp001}, we determine the sign of the ZM Weyl invariants $I_1$ and $K_2$, i.e. the regions in the $r-\theta$ space with negative and positive values for these two invariants, for various values of the Kerr parameter $a$, electric charge $q$ and acceleration $\alpha$ of the black hole. We also determine the regions of negative and positive values, in the $r-\theta$ space for the mixed ZM curvature invariant $I_9$ in Figure \ref{regionmI9a09939q011alp001}, for the set of the black hole parameters $a=0.9939,q=0.11,\alpha=0.01,m=1$.
Contour plots in the $r-\theta$ space, for the ZM invariants $I_1, K_2$, are displayed in Figures \ref{LevelcurvI1a09939q011al001}-\ref{LevelcurvK2a09939q011al001}, for a high spin value $a=0.9939$, electric charge $q=0.11$ and acceleration parameter $\alpha=0.01$.
In Figure \ref{DreiGraphChernPontryaginK2accel}, armed with our explicit solution for the Chern-Pontryagin ZM invariant, eqn. (\ref{CPBeschleunigungKNdS}), we display in 3d plots, the Hirzebruch density as a function  of the Boyer-Lindquist coordinates $r$ and $\theta$ for various sets of values for the Kerr parameter $a$,electric charge $q$ for an accelerating charged and rotating black hole of mass $m=1$ and fixed value of the acceleration parameter $\alpha=0.05$.
In Figures \ref{diag3dI9I10K2a052q04alp001},\ref{diag3dI1K2I9I10a09939q011alp001} we plot
the Weyl invariants $I_1,K_2$ and the mixed invariants $I_9,I_{10}$ as a function of the Boyer-Lindquist coordinates $r$ and $\theta$ for an accelerating, rotating and charged black of mass $m=1$, acceleration $\alpha=0.01$ for two different sets of values for its Kerr parameter $a$ and electric charge $q$.

We now briefly comment the meaning/importance of regions with positive and negative values of the invariants $I_1,K_2$ and the zero-value boundaries between the regions in the graphs.
The regions of spacetime where the invariants $I_1,K_2$ vanish can be determined analytically. For reasons of simplicity of the presentation we focus the discussion in the case of zero acceleration, i.e. $\alpha=0$.
Solving  $I_1=0$ for $\cos(\theta)$, applying the method of Tartaglia and Cardano, and assuming $\alpha=0$ we obtain:
\begin{align}
&\cos(\theta)=\pm \frac{a^2 q^2-3 a^2 m r}{3 a^3 m}\nonumber \\
&\pm\left(-a^4 q^4+12 a^4 m q^2 r-18 a^4 m^2 r^2\right)/\Biggl(3 a^3
m \Biggl(-a^6 q^6+18 a^6 m q^4 r-54 a^6 m^2 q^2 r^2+54 a^6 m^3 r^3\nonumber \\
&+3 \sqrt{6} \sqrt{-a^{12} m^2 q^8 r^2+18 a^{12} m^3 q^6 r^3-72 a^{12} m^4 q^4 r^4+108
a^{12} m^5 q^2 r^5-54 a^{12} m^6 r^6}\Biggr)^{1/3}\Biggr)\nonumber \\
&\mp\frac{1}{3 a^3 m}\Biggl(-a^6 q^6+18 a^6 m q^4 r-54 a^6 m^2 q^2 r^2+54 a^6 m^3 r^3\nonumber \\
&+3 \sqrt{6}
\sqrt{-a^{12} m^2 q^8 r^2+18 a^{12} m^3 q^6 r^3-72 a^{12} m^4 q^4 r^4+108 a^{12} m^5 q^2 r^5-54 a^{12} m^6 r^6}\Biggr)^{1/3},\label{synoro1I1} \\
&\cos(\theta)=\pm\frac{a^2 q^2-3 a^2 m r}{3 a^3 m}\nonumber \\
&\mp\left(\left(1+i \sqrt{3}\right) \left(-a^4 q^4+12 a^4 m q^2 r-18 a^4 m^2 r^2\right)\right)/\Biggl(6 a^3 m \Biggl(-a^6
q^6+18 a^6 m q^4 r-54 a^6 m^2 q^2 r^2+54 a^6 m^3 r^3\nonumber \\
&+3 \sqrt{6} \sqrt{-a^{12} m^2 q^8 r^2+18 a^{12} m^3 q^6 r^3-72 a^{12} m^4 q^4 r^4+108 a^{12}
m^5 q^2 r^5-54 a^{12} m^6 r^6}\Biggr)^{1/3}\Biggr)\nonumber \\
&\pm\frac{1}{6 a^3 m}\left(1-i \sqrt{3}\right) \Biggl(-a^6 q^6+18 a^6 m q^4 r-54 a^6 m^2 q^2 r^2+54
a^6 m^3 r^3\nonumber \\
&+3 \sqrt{6} \sqrt{-a^{12} m^2 q^8 r^2+18 a^{12} m^3 q^6 r^3-72 a^{12} m^4 q^4 r^4+108 a^{12} m^5 q^2 r^5-54 a^{12} m^6 r^6}\Biggr)^{1/3},
\end{align}
\begin{align}
&\cos(\theta)=\pm\frac{a^2 q^2-3 a^2 m r}{3 a^3 m}\nonumber \\
&\mp\left(\left(1-i \sqrt{3}\right) \left(-a^4 q^4+12 a^4 m q^2 r-18 a^4 m^2 r^2\right)\right)/\Biggl(6 a^3 m \Biggl(-a^6
q^6+18 a^6 m q^4 r-54 a^6 m^2 q^2 r^2+54 a^6 m^3 r^3\nonumber \\
&+3 \sqrt{6} \sqrt{-a^{12} m^2 q^8 r^2+18 a^{12} m^3 q^6 r^3-72 a^{12} m^4 q^4 r^4+108 a^{12}
m^5 q^2 r^5-54 a^{12} m^6 r^6}\Biggr)^{1/3}\Biggr)\nonumber \\
&\pm\frac{1}{6 a^3 m}\left(1+i \sqrt{3}\right) \Biggl(-a^6 q^6+18 a^6 m q^4 r-54 a^6 m^2 q^2 r^2+54
a^6 m^3 r^3\nonumber \\
&+3 \sqrt{6} \sqrt{-a^{12} m^2 q^8 r^2+18 a^{12} m^3 q^6 r^3-72 a^{12} m^4 q^4 r^4+108 a^{12} m^5 q^2 r^5-54 a^{12} m^6 r^6}\Biggr)^{1/3}\label{synoro3I1},
\end{align}
while solving $K_2=0$ (again for zero acceleration $\alpha=0$) yields:
\begin{align}
\cos(\theta)&=0,\label{Hirzebruchroot1}\\
\cos(\theta)&=\pm\sqrt{\frac{3m r^2-2q^2r}{m}}\frac{1}{a},\\
\cos(\theta)&=\pm\frac{1}{a}\sqrt{\frac{q^2r^2-mr^3}{q^2-3mr}}\label{Hirzebruchdreiroot}.
\end{align}

Thus the zero boundary expressed by eqns(\ref{synoro1I1})-(\ref{synoro3I1}) can be interpreted as separating regions of electric dominance of the Weyl tensor ($I_1>0$) from regions of Weyl magnetic dominance ($I_1<0$).
We mention at this point that an observer with a timelike velocity
vector field $u^{\alpha}$ is said to measure the \textit{electric} and \textit{magnetic} components, $\mathcal{E}_{\alpha\beta}$ and $\mathcal{H}_{\alpha\beta}$, respectively, of the Weyl tensor by
\begin{equation}
\mathcal{E}_{\alpha\beta}\equiv C_{\alpha\gamma\beta\delta}u^{\gamma}u^{\delta}
,\;\;\;\mathcal{H}_{\alpha\beta}\equiv C^*_{\alpha\gamma\beta\delta}u^{\gamma}u^{\delta}.
\end{equation}
The curvature invariant $I_1$ is related to the electric and magnetic components of the Weyl tensor as follows \cite{Filipe}:
\begin{equation}
\frac{I_1}{8}=\mathcal{E}^{\alpha\beta}\mathcal{E}_{\alpha\beta}-\mathcal{H}^{\alpha\beta}\mathcal{H}_{\alpha\beta},
\label{dominanceWeylEM}
\end{equation}
while the Chern-Pontryagin invariant $K_2$ is expressed as follows \cite{Filipe}:
\begin{equation}
\frac{1}{16}K_2=\mathcal{E}^{\alpha\beta}\mathcal{H}_{\alpha\beta}.
\end{equation}
Equation (\ref{dominanceWeylEM}) clarifies the introduction of the region of Weyl electric dominance: $\mathcal{E}^{\alpha\beta}\mathcal{E}_{\alpha\beta}>\mathcal{H}^{\alpha\beta}\mathcal{H}_{\alpha\beta}$, i.e. $I_1>0$, and regions of Weyl magnetic dominance: $\mathcal{E}^{\alpha\beta}\mathcal{E}_{\alpha\beta}<\mathcal{H}^{\alpha\beta}\mathcal{H}_{\alpha\beta}$, i.e. $I_1<$0.

The zeros of the Hirzebruch invariant $K_2$  in eqns.(\ref{Hirzebruchroot1})-(\ref{Hirzebruchdreiroot}) define purely electric/magnetic Weyl surfaces. For $\theta=\pi/2$ and $\cos(\theta)=\pm\frac{\sqrt{mr(3mr-2q^2)}}{ma}$ Weyl tensor is purely electric while for the zeros in  (\ref{Hirzebruchdreiroot}) the Weyl tensor is purely magnetic.
A Weyl tensor is called purely electric (purely magnetic) when $\mathcal{H}_{\alpha\beta}=0$ ($\mathcal{E}_{\alpha\beta}=0$) \cite{arianhod}.

For the vacuum Kerr spacetime ($q=\alpha=\Lambda=0$), $K=I_1$,  the zeros in Eqns.(\ref{synoro1I1})-(\ref{synoro3I1}) reduce to $r=\pm a \cos(\theta)$ and $r=\pm(2\pm\sqrt{3})a\cos(\theta)$ and signal transitions between regions of gravitoelectric ($K>0$) vs gravitomagnetic ($K<0$) dominance \cite{cbcruf}.
The zeros of the Hirzebruch invariant $K_2$ in eqns.(\ref{Hirzebruchroot1})-(\ref{Hirzebruchdreiroot}) for $q=0$ define purely electric/magnetic surfaces and occur for $\theta=\pi/2$ and $r=\pm \frac{a\cos(\theta)}{\sqrt{3}}$ (purely electric) and $r=\pm a \sqrt{3}\cos(\theta)$
(purely magnetic).
We note that the spatial fields associated with a measurement of the Riemann tensor by an arbitrary observe $O(\mathbf{u})$ are defined by \cite{Filipe},\cite{cbcruf}:
\begin{align}
\mathbb{E}_{\alpha\beta}\equiv &R_{\alpha\mu\beta\nu}u^{\mu}u^{\nu},\;\;\;\mathbb{H}_{\alpha\beta}\equiv R^{*}_{\alpha\mu\beta\nu}u^{\mu}u^{\nu},\\
&\mathbb{F}_{\alpha\beta}\equiv ^*R^{\;*}_{\alpha\mu\beta\nu}u^{\mu}u^{\nu}.
\end{align}
The Kretschmann and the Chern-Pontryagin scalars are gives in terms of these tensors by:
\begin{align}
\frac{1}{4}K&=\mathbb{E}^{\alpha\gamma}\mathbb{E}_{\alpha\gamma}+\mathbb{F}^{\alpha\gamma}\mathbb{F}_{\alpha\gamma}
-2\mathbb{H}^{\alpha\gamma}\mathbb{H}_{\alpha\gamma},\\
\frac{1}{8}K_2&=(\mathbb{E}^{\alpha\gamma}-\mathbb{F}^{\alpha\gamma})\mathbb{H}_{\alpha\gamma}
\end{align}

Analogous definitions as those given above for the \textit{purely Weyl magnetic} and \textit{purely Weyl electric} tensors can be provided for a nonvacuum Riemann tensor. In particular, the Riemann tensor and the spacetime is called \textit{purely electric} (\textit{purely magnetic}) at a point if an observer  $O(\mathbf{u^{\prime}})$ exists for which $ \mathbb{H}_{\alpha\beta}^{\prime}=0$ ($\mathbb{E}_{\alpha\beta}^{\prime}=0$) and moreover $\mathbb{E}_{\alpha\beta}^{\prime}\not=0$ ($ \mathbb{H}_{\alpha\beta}^{\prime}\not=0$) for \textit{all} observers \footnote{The reason for the more refined definitions for the Riemann tensor is that for a nonvacuum Riemann tensor the conditions $\mathbb{E}_{\alpha\beta}^{\prime}=0$ and $ \mathbb{H}_{\alpha\beta}^{\prime}=0$ may hold simultaneously even for the same observer $O(\mathbf{u^{\prime}})$ \cite{haddow}. } $O(\mathbf{u^{\prime}})$ \cite{arianhod},\cite{Filipe}.
One can then investigate interesting questions such as developing criteria for the conditions $\mathbb{E}_{\alpha\beta}^{\prime}=0$ or $ \mathbb{H}_{\alpha\beta}^{\prime}=0$, further physical interpretations for all the ZM invariants and questions relating to frame-dragging, gyroscopic forces, geodesic deviation  for accelerating rotating and charged black holes with $\Lambda\not =0$. A systematic investigation of these issues is however beyond the scope of the current work and it will be a subject of a future publication.

\section{Discussion and Conclusions}\label{symperasmata}

In this work we computed, exact algebraic expressions for the curvature invariants in the ZM framework, for two of the most general black hole solutions. Namely, i)  for the Kerr-Newman-(anti-)de Sitter black hole metric ii) for accelerating Kerr-Newman black hole in (anti-)de Sitter spacetime.
Despite the complexity of the computations involved using the tensorial method of calculation, our final expressions are reasonably compact and easy to use in applications.
Our results are culminated in Theorems \ref{courbureinvriemein} and \ref{courbureinvtheoriazwei}.
We also obtained explicit algebraic expressions for the Kretschmann invariant for both the Kerr-Newman-(anti-)de Sitter black hole geometry and for accelerating charged and rotating black holes with $\Lambda\not=0$, Theorems \ref{KretschmannKNdS} and \ref{KretschmannAccelaKNdS} respectively. Furthermore, we computed novel explicit algebraic expressions for the topological Euler-Poincare invariant $K_{{\rm Euler}}$, see Eqn.(\ref{LeonardEuler}) for the case of a non-accelerating KN(a-)dS black hole \footnote{We have also computed the explicit expression for the Euler-Poincare invariant for accelerating KN(a-)dS black holes. However, the resulting expression is lengthy and cumbersome to reproduce it here. Nevertheless we have checked that it satisfies the fundamental identities in (\ref{avezmacaroni}) that resulted from Avez's theorem. This means that $K_{{\rm Euler}}^{accel-KN(a)dS}=-(\ref{AcceBeschlKNdSein})+2(\ref{riccisquare})-\frac{2}{3}(\ref{constcurvaaccel})^2.$}.
We have also checked our computations through the NP formalism and with the discovery of certain syzygies that the curvature invariants satisfy-see Eqns(\ref{syzyg1}),(\ref{syzygzwei}),(\ref{syzdrei}),(\ref{syzvier}).

From the bestiary of our explicit novel expressions for the ZM curvature invariants, we performed an extensive plotting  of curvature that represents a novel pathway to explore  the geometry of  spacetime inside accelerating and non-accelerating Kerr-Newman black holes in (anti-)de Sitter spacetime. In the process we demarcated the regions in the $r-\theta$ space of negative and positive sign for these curvature invariants.

Interesting potential applications, among others, include: the Chern-Pontryagin invariant-eqn(\ref{Hirzebruch}), also appears to play a role in the electromagnetic duality anomaly in curved spacetimes. As is known the source-free Maxwell action is invariant under electric-magnetic duality rotations \footnote{$F_{\mu\nu}\rightarrow F_{\mu\nu}\cos(\theta)+F^{*}_{\mu\nu}\sin(\theta).$} in arbitrary spacetimes \cite{desboim}. This leads to a conserved classical Noether charge. In \cite{agulrionavsa},inspired by earlier work \cite{dolgovkvza}\footnote{The result of  \cite{dolgovkvza}, was further explored in \cite{praktoreut} where it was shown that for antisymmetric gauge fields of rank $2n-1$ coupled to gravity in $4n$ dimensions, the symmetry under duality rotations is broken by quantum effects.}, it was shown that this conservation law was broken at the quantum level in the presence of a background field with a non-trivial Chern-Pontryagin invariant.

In particular quantum effects may induce violation of helicity conservation for photons propagating in curved spacetimes. Observable effects of this photon chiral anomaly are directly related to the variation of electromagnetic helicity $\mathcal{H}_{\rm em}$ \cite{GalaGabriel}:

\begin{equation}
\Delta\langle\mathcal{H}_{\rm em}\rangle\propto\int_{t_1}^{t_2}\int_{\Sigma^3}R_{\alpha\beta\mu\nu}R^{*\alpha\beta\mu\nu}\sqrt{-g}{\rm d}^4 x
\end{equation}
If the integral on the right term is different from zero, then  $\mathcal{H}_{\rm em}$ is not conserved. The difference between the numbers of right circularly polarised photons and left  circularly polarised photons changes: the degree of circular polarisation is not conserved.
Indeed, at large distances from a Kerr-Newman-(anti-)de Sitter black hole the Hirzebruch density, eqn.(\ref{CHERNPOTRYAGINNOBESCH}), has the expansion:
\begin{align}
K_2&=^*\mathbf{R}\cdot\mathbf{R}=\frac{96a}{r^{12}}\Biggl(-3m^2 r^5\cos(\theta)+5 \cos(\theta) m q^2 r^4-2 q^4 r^3\cos(\theta)\nonumber \\
&+10 m^2 a^2 r^3 \cos^3(\theta)\Biggr)\left(1-\frac{6a^2}{r^2}\cos^2(\theta)+\cdots\right)\nonumber \\
&=-288 \frac{m^2 a}{r^7}\cos(\theta)+480\frac{a\cos(\theta)mq^2}{r^8}\nonumber \\
&-192 \frac{a\cos(\theta)}{r^9}\left(q^4-14 m^2 a^2 \cos^2(\theta)\right)+O\left(\frac{1}{r^9}\right)
\end{align}
Then integration yields the result:
\begin{align}
\int R_{\alpha\beta\mu\nu}R^{*\alpha\beta\mu\nu}\sqrt{-g}{\rm d}^4 x\propto
\int_0^{\pi}\cos(\theta) \sin^2(\theta) (r^2+a^2\cos^2(\theta)){\rm d}\theta=0.
\label{nullintegraCPH}
\end{align}
Thus despite the fact that, for a non-accelerating KN(a)dS black hole, the Hirzebruch invariant is non-trivial its integral over all space is zero-in this case there are no observable effects related to the quantum anomaly. This is in agreement with the recent calculation for the Kerr metric in \cite{GalaGabriel}.

Let us investigate now the case of accelerating Kerr-Newman black holes in (anti-)de Sitter spacetime. The Chern-Pontryagin invariant $K_2$ in equation (\ref{CPBeschleunigungKNdS}) for large radii takes the form:
\begin{align}
K_2&=^*\mathbf{R}\cdot\mathbf{R}=\frac{96a\left(\alpha r \cos \! \left(\theta \right)-1\right)^{6}}{r^{12}}\Biggl(\cos \! \left(\theta \right)^{3}a^{4}\alpha m +\cos \! \left(\theta \right)^{3}a^{2}\alpha q^{2}r \nonumber \\
&-3\cos \! \left(\theta \right)a^{2}\alpha m \,r^{2}-\cos \! \left(\theta \right)\alpha q^{2}r^{3}-3a^{2}\cos \! \left(\theta \right)^{2}m r +a^{2}\cos \! \left(\theta \right)^{2}q^{2}+m \,r^{3}-q^{2}r^{2}\Biggr) \nonumber \\
 &\times \Biggl[3\cos \! \left(\theta \right)^{2}a^{2}\alpha m r +2\cos \! \left(\theta \right)^{2}\alpha q^{2}r^{2}+\cos \! \left(\theta \right)^{3}a^{2}m -\alpha r^{3}m -3\cos \! \left(\theta \right)m \,r^{2} \nonumber \\
 &+2\cos \! \left(\theta \right)q^{2}r\Biggr]\left(1-6\frac{a^2}{r^2}\cos^2(\theta)+\cdots\right)\nonumber \\
 &=\frac{96a\left(\alpha r \cos \! \left(\theta \right)-1\right)^{6}}{r^{6}}(mq^2\alpha^2\cos(\theta)-m^2\alpha)+\cdots
\label{GVKradiation}
\end{align}
Interestingly, the following  polar angular integration  of the leading term in (\ref{GVKradiation}), which is a part of the spacetime integral $\iiiint R_{\alpha\beta\mu\nu}R^{*\alpha\beta\mu\nu}\sqrt{-g}{\rm d}^4 x$, gives a non-zero result:
\begin{align}
\int_0^\pi\frac{96a}{r^6}(mq^2\alpha^2\cos(\theta)-m^2\alpha)\sin^2(\theta)(r^2+a^2\cos^2(\theta))(\alpha r\cos(\theta)-1)^2{{\rm d}\theta}\not =0.
\label{nonnullCPHintegraGVK}
\end{align}
Thus, it appears that accelerating Kerr-Newman black holes in (anti-)de Sitter spacetime  yield a non-zero effect for the quantum photon chiral anomaly since a nonzero Chern-Pontryagin integrated term is present.

A more detailed investigation of  the possible non-null effect for the quantum chiral anomaly of circularly polarised photons for accelerating, rotating and charged black holes with $\Lambda\not =0$ and its likely observational consequences is beyond the scope of the current paper.
It is certainly worth exploring this issue of quantum chiral anomaly in the context of accelerating Kerr-Newman black holes with $\Lambda\not=0$ further in a separate publication \cite{KraniotisGVchiral}, for the Hirzebruch signature density in eqn.(\ref{CPBeschleunigungKNdS}).

Another interesting and related application would be to consider a scalar field coupled to the divergence of the photonic chiral current, the Hirzebruch signature density then acts as a source for the axion, and investigate for the Kerr-Newman-(anti-)de Sitter background. Therefore, generalising the work in \cite{MR} for Kerr black holes.

Last but not least, we aim to generalise our work on gravitational lensing and the black hole shadow in Kerr-Newman and Kerr-Newman-(anti-)de Sitter spacetimes  \cite{GRGKRANIOTIS},\cite{GVKraniotisCQG2011} to the more involved case of accelerating KN(a)dS black holes \cite{GVKraniotisAccelBH}.

\begin{acknowledgement}
I thank Dr. George Poulipoulis and Dr. S. Skopoulis for useful discussions.
\end{acknowledgement}


\begin{thebibliography}{99}

\bibitem{Zahkary} E. Zakhary and. C. B. G. McIntosh, \textit{A Complete Set of Riemann Invariants}, Gen.Rel.Grav.29 (1997),539-581

    \bibitem{GeheDebr} J. G\'{e}h\'{e}niau and R. Debrever, \textit{Les quatorze invariants de courbure de l'espace riemannien \'{a} quatre dimensions},Helvetica Physica Acta 29,(1956),101-105

    \bibitem{LWitten} L. Witten, \textit{Invariants of General Relativity and the Classification of Spaces}, Phys.Rev. 113 (1959),pp 357-362
        \bibitem{HarveyAL} A. Harvey,\textit{On the algebraic invariants of the four-dimensional Riemann tensor},Class.Quan.Grav 7(1990),715-716
             \bibitem{CarminatiMcL} J. Carminati and R. G. McLenaghan,\textit{Algebraic invariants of the Riemann tensor in a four-dimensional Lorentzian space}, J. Math.Phys.\textbf{32}, (1991) pp 3135-3140
    \bibitem{Petrov} A Z Petrov, \textit{The Classification of Spaces Defining Gravitational Fields}, Gen.Rel.Grav.\textbf{32} (2000),1665-1685, Original title: Klassifikacya prostranstv opredelyayushchikh polya tyagoteniya. Uchenye Zapiski
Kazanskogo Gosudarstvennogo Universiteta im. V. I. Ulyanova-Lenina [Scientific Proceedings
of Kazan State University, named after V.I. Ulyanov-Lenin], 114 (8), 55–69 (1954).



    \bibitem{Ciufolini} I. Ciufolini, \textit{Dragging of Inertial Frames, Gravitomagnetism, and Mach's Principle},in Einstein Studies 6,Mach's Principle, From Newton's Bucket to Quantum Gravity,Birkh\"{a}user,(1995),pp 386-402
        \bibitem{BakerJMC} J. Baker and M. Campanelli, \textit{Making use of geometrical invariants in black hole collisions}, Phys.Rev.D 62 (2000) 127501
            \bibitem{Filipe} L. Filipe \textit{et al},\textit{Gravitomagnetism and the significance of the curvature scalar invariants},Phys.Rev.D 104(2021)084081

        \bibitem{Kretschmann} E. Kretschmann, \textit{\"Uber den physikalischen Sinn der Relativit\"atspostulate. A. Einsteins neue und seine urspr\"ungliche Relativit\"atstheorie},Annalen der Physik \textbf{53},(1917) 575-614

\bibitem{GrifPod} J. B. Griffiths and Ji\v{r}\'{\i} Podolsk\'{y},
              \textit{Exact spacetimes in Einstein's General Relativity, }Cambridge
            Monographs on Mathematical Physics, Cambirdge University Press (2009)


    \bibitem{PlebanskiDemianski}J. F. Plebanski, M. Demianski, \textit{Rotating, Charged, and Uniformly Accelerating Mass in General Relativity}, Annals of Physics \textbf{98},(1976) 98-127
\bibitem{Supern} S. Perlmutter \textit{et al}, Astrophys.Journal
\textbf{517}(1999) 565; A. V. Filippenko \textit{et al}
Astron.J.\textbf{116} 1009

\bibitem{Jones} D. O. Jones \textit{et al}, \textit{The Foundation Supernova Survey: Measuring Cosmological Parameters with Supernovae from a Single Telescope}  Astrophys. J. \textbf{881}, (2019) 19
\bibitem{Aubourg} E. Aubourg \textit{et al}, \textit{Cosmological implications of baryon acoustic oscillation measurements}, Phys.Rev.D \textbf{92}, (2015) 123516
\bibitem{AbbottTMC} T.M.C. Abbott \textit{et al},\textit{Dark Energy Survey Year 3 results: Cosmological constraints from galaxy clustering and weak lensing}, Phys.Rev.D \textbf{105}, (2022) 023520

    \bibitem{GVKSWB} G. V. Kraniotis and S. B. Whitehouse, \textit{General relativity, the
cosmological constant and modular forms} Class. Quantum Grav.
\textbf{19} (2002), 5073-5100


\bibitem{Zajacek} M. Zaja$\breve{\rm c}$ek,  A.Tursunov, A. Eckart and S. Britzen, \textit{On the charge of the Galactic centre black hole}, Mon.Not.Roy.Astron.Soc. \textbf{480}, (2018) 4408-4423

    \bibitem{Britzen} A. Tursunov, M Zaja$\breve{\rm c}$ek, A. Eckart, M. Kolos, S. Britzen, Z. Stuchl\'{\i}k, B. Czerny, and V. Karas, \textit{Effect of Electromagnetic Interaction on Galactic Center Flare Components}, Astrophys.J. \textbf{897} (2020) 1, 99
\bibitem{universemdpi} Z. Stuchl\'{\i}k, M. Kolo$\breve{\rm s}$,
J. Kov\'{a}$\breve{\rm r}$, P. Slan\'{y}, and A. Tursunov, \textit{Influence of Cosmic Repulsion and Magnetic Fields on Accretion Disks Rotating around Kerr Black Holes}, Universe 6 (2020) 2, 26

\bibitem{waldcharge} A. Tursunov, Z. Stuchl\'{\i}k, M. Kolo$\breve{\rm s}$, N. Dadlich, and B. Ahmedov, \textit{Supermassive Black Holes as Possible Sources of Ultrahigh-energy Cosmic Rays},  Astrophys.J. \textbf{895} (2020) 1, 14


  \bibitem{HenryRC} R.C. Henry,\textit{Kretschmann scalar for a Kerr-Newman black hole}, The Astrophys.J.\textbf{535}, (2000) 350-353

       \bibitem{LakeK} K. Lake,\textit{Invariants of the Kerr Vacuum}, Gen.Rel.Grav.35 (2003),2271-2277

\bibitem{PMusKLak} P. Musgrave and K. Lake, \textit{Scalar invariants of the Kerr-Newman metric: a simple application of GRTensor}, Comp.in Phys.8 (1994),589


    \bibitem{cbcruf} C. Cherubini, D. Bini, S. Capozziello and R. Ruffini, \textit{Second order scalar invariants of the Riemann tensor:application to black hole spacetimes},Int.J.Mod.Phys.D 11(2002) 827


        \bibitem{overduincwh} J. Overduin, M. Coplan, K. Wilcomb and R. C. Henry,\textit{Curvature invariants for charged and rotating black holes}, Universe (2020),6,22

            \bibitem{mattinglyCleav} Mattingly B.; Kar A.;Gorban M.; Julius W.;Watson C.K.;Ali M.D.;Baas A.;Elmore C.;Lee J.S.;Shakerin B.;Davis E.W.;Cleaver G.B. Curvature Invariants for the Accelerating Nat\'{a}rio Warp Drive. {\em Particles} {\bf 2020},3,642
                \bibitem{GeraldC} Mattingly B.; Kar A.;Gorban M.; Julius W.;Watson C.K.;Ali M.D.;Baas A.;Elmore C.;Lee J.S.;Shakerin B.;Davis E.W.;Cleaver G.B. Curvature Invariants for the
               Alcubierre and Nat\'{a}rio Warp Drives. {\em Universe} {\bf 2021},7,21
                \bibitem{Boos} Boos J. Pleba\'{n}ski-Demia\'{n}ski solution of general relatvity and its expressions quadratic and cubic in curvature: Analogies to electromagnetism. {\em  Int.J.Mod.Phys.D} {\bf 2015},24,1550079
        \bibitem{mccallumr} M.A.H. MacCallum, \textit{Computer algebra in gravity research}, Liv.Rev.Relat.(2018) 21:6
            \bibitem {Newman} Newman E. T.;  Couch E.; Chinnapared K.; Exton  A.; Prakash A.;
 Torrence R. Metric of a Rotating, Charged Mass. {\em J. Math. Phys.} {\bf  1965},6,918
\bibitem {KerrR} Kerr R. P. Gravitational field of a spinning mass as an
example of algebraically special metrics. {\em Phys. Rev. Lett.} {\bf 1963},11,237
\bibitem {Stuchlik1}Z. Stuchl\'{\i}k, G. Bao, E. \O stgaard and S.
Hled\'{\i}k,\textit{\ Kerr-Newman-de Sitter black holes with a restricted
repulsive barrier of equatorial photon motion, }Phys. Rev. D. \textbf{58}
(1998) 084003
 \bibitem {BCAR}B. Carter, \textit{Global structure of the Kerr family of
gravitational fields }Phys.Rev.\textbf{174 }(1968)1559-71


\bibitem{ZdeStu} Z. Stuchl\'{\i}k and S.Hled\'{\i}k,
\textit{Equatorial photon motion in the Kerr-Newman spacetimes
with a non-zero cosmological constant}, Class. Quantum Grav.
\textbf{17} (2000) 4541-4576

         \bibitem{ZST} Z. Stuchl\'{\i}k, \textit{The motion of test particles in black-hole backgrounds
           with non-zero cosmological constant}, Bull. of the Astronomical
             Institute of Chechoslovakia \textbf{34} (1983) 129-149


\bibitem{szekeres} P. Szekeres, \textit{The Gravitational Compass}, J.Math.Phys. 6 (1965) 1387

\bibitem{Dianyan} X. Dianyan,\textit{Two important invariant identities} Phys.Rev.D 35,(1987),pp 769-770
    \bibitem{Avez} A. Avez, \textit{Characteristic Classes and Weyl Tensor: Applications to General Relativity}, Proc.Nat.Acad.Sciences, 66, (1970), 265-268
        \bibitem{duff} M.J. Duff, \textit{Weyl,Pontryagin, Euler, Eguchi and Freund}, J. Phys.A.Math. Theor. \textbf{53} 2020, 301001
            \bibitem{solodukhin} A.F. Astaneh and S.N. Solodukhin, Phys.Lett.B 816 (2021) 136282
                \bibitem{genzelr} R. Genzel \textit{et al}, \textit{Near-infrared flares from accreting gas around the supermassive black hole at the Galactic
Centre} Nature 425 (2003) 934
\bibitem{aschenbach} B. Aschenbach \textit{et al}, \textit{X-ray flares reveal mass and angular momentum of
the Galactic Centre black hole}, Astron. Astrophys.\textbf{417} (2004), 71-78

    \bibitem{EHT} The Event Horizon Telescope Collaboration, \textit{First M87 Event Horizon Telescope Results. I. The Shadow of the Supermassive Black Hole}, The Astrophysical Journal Letters, 875: L1 (2019) April 10
        \bibitem{Eisenhauer} F. Eisenhauer \textit{et al,} \textit{Sinfoni in the
galactic centre: young stars and \ infrared flares in the central light-month
}(2005) Astrophys.J. \textbf{628} 246-59
\bibitem{schmidt} H. -J. Schmidt, {\em The Square of the Weyl Tensor can be Negative}, Gen.Relativ.Grav.35 (2003) 937-938
\bibitem{NPformalismCI} Newman E.;   Penrose R. An Approach to Gravitational Radiation by a Method of Spin Coefficients. {\em J. Math.Phys.} {\bf 1962 },3,566
\bibitem{KraniotisDirac} G. V. Kraniotis, \textit{The massive Dirac equation in the Kerr-Newman-de Sitter and Kerr-Newman black hole spacetimes} J.Phys.Comm. 3 (2019) 035026,[arXiv:1801.03157 ]
        \bibitem{Chandrasekhar} S. Chandrasekhar, \textit{The Mathematical Theory of Black Holes} Oxford Classic Texts in Physical Sciences, 1992
            \bibitem{spinors} R. Penrose and W. Rindler, \textit{Spinors and space-time} Vol.I: \textit{Two-spinor calculus and relativistic fields} Cambridge Monographs in Mathematical Physics,(1984) CUP; Vol.2,\textit{Spinor and twistor methods in space-time geometry} (1986) CUP

        \bibitem{PodolskyGrif} J. Podolsk\'{y} and J. B. Griffiths, \textit{Acccelerating Kerr-Newman black holes in (anti)-de Sitter spacetime}, Phys.Rev.D 73 (2006),044018

            \bibitem{arneGrezen} A. Grenzebach, V. Perlick and C. L\"ammerzahl,\textit{Photon regions and Shadows of accelerated black holes}, Int.J.Mod.Phys.D 24 (2015) 09, 1542024
                \bibitem{arianhod} R. Arianhod \textit{et al}, \textit{Magnetic curvatures} Class. Quantum Grav.\textbf{11} (1994) 2331-2335
                    \bibitem{haddow} B. M. Haddow, \textit{Purely magnetic space-times}, J. Math.Phys.\textbf{36}(1995)5848
        \bibitem{desboim} S. Deser and C. Teitelboim, Duality transformations of Abelian and non-Abelian gauge fields, Phys.Rev.D 13(1976),1592-1597
\bibitem{agulrionavsa} I Agullo, A. del Rio and J. Navarro-Salas,\textit{Electromagnetic duality anomaly in curved spacetime}, Phys.Rev.Let.118 (2017)111301
        \bibitem{dolgovkvza} A.D. Dolgov, I.B. Khriplovich,   A.I. Vainshtein, V.I. Zakharov \textit{Photonic chiral current and its anomaly in a gravitational field}, Nucl.Phys.B315(1989) 138-152
 \bibitem{praktoreut} M. Reuter, \textit{Chiral anomaly of antisymmetric tensor fields},Phys.Rev.D 37(1988),1456-1463
     \bibitem{GalaGabriel} M. Galaverni and G. S.J. Gabriele,\textit{Photon helicity and quantum anomalies in curved spacetimes}, Gen.Rel.Gravit. (2021)53:46
         \bibitem{KraniotisGVchiral} G. V. Kraniotis, Work in Progress
         \bibitem{MR} M. Reuter, \textit{A mechanism generating axion hair for Kerr black holes}, Class.Quantum Grav 9 (1992) 751-756
             \bibitem{GRGKRANIOTIS} G. V. Kraniotis, \textit{Gravitational lensing and frame dragging of light in the Kerr-Newman and the Kerr-Newman-(anti) de Sitter black hole spacetimes}, Gen. Rel. Grav. {\bf 46} (2014) 1818
                  \bibitem{GVKraniotisCQG2011} G. V. Kraniotis \textit{Precise analytic treatment of Kerr and Kerr-(anti) de Sitter black holes as gravitational lenses}, Class. Quant.Grav. {\bf 28} (2011) 085021
                 \bibitem{GVKraniotisAccelBH}  G. V. Kraniotis, Work in Progress
    \end{thebibliography}
 \end{document}